\documentclass[12pt]{iopart}

\usepackage[dvips]{graphicx}
\usepackage{bbm}
\usepackage{psfrag}
\usepackage{iopams}

\newcommand{\R}{\mathbbm{R}}
\newcommand{\N}{\mathbbm{N}}
\newcommand{\Z}{\mathbbm{Z}}
\newcommand{\tfrac}[2]{{\textstyle \frac{#1}{#2}}}

\newtheorem{theorem}{Theorem}

\begin{document}

\title{Bifurcations of discrete breathers in a diatomic Fermi-Pasta-Ulam chain}

\author{Guillaume James$^1$ and Michael Kastner$^2$}
\address{$^1$ Laboratoire Math\'ematiques pour l'Industrie et la Physique (UMR 5640), INSA de Toulouse, 135 avenue de Rangueil, 31077 Toulouse Cedex 4, France}
\address{$^2$ Physikalisches Institut, Lehrstuhl f\"ur Theoretische Physik I, Universit\"at Bayreuth, 95440 Bayreuth, Germany}
\ead{\mailto{Guillaume.James@insa-toulouse.fr}, \mailto{Michael.Kastner@uni-bayreuth.de}}
\date{\today}

\begin{abstract}
Discrete breathers are time-periodic, spatially localized solutions of the equations of motion for a system of classical degrees of freedom interacting on a lattice. Such solutions are investigated for a diatomic Fermi-Pasta-Ulam chain, i.\,e., a chain of alternate heavy and light masses coupled by anharmonic forces. For hard interaction potentials, discrete breathers in this model are known to exist either as ``optic breathers'' with frequencies above the optic band, or as ``acoustic breathers'' with frequencies in the gap between the acoustic and the optic band. In this paper, bifurcations between different types of discrete breathers are found numerically, with the mass ratio $m$ and the breather frequency $\omega$ as bifurcation parameters. We identify a period tripling bifurcation around optic breathers, which leads to new breather solutions with frequencies in the gap, and a second local bifurcation around acoustic breathers. These results provide new breather solutions of the FPU system which interpolate between the classical acoustic and optic modes. The two bifurcation lines originate from a particular ``corner'' in parameter space $\left(\omega,m\right)$. As parameters lie near this corner, we prove by means of a center manifold reduction that small amplitude solutions can be described by a four-dimensional reversible map. This allows us to derive formally a continuum limit differential equation which characterizes at leading order the numerically observed bifurcations.
\end{abstract}

\pacs{02.30.Oz, 05.45.-a, 45.05.+x, 63.20.Pw}

\section{Introduction}

Discrete breathers are time-periodic, spatially localized solutions of the equations of motion for a system of classical degrees of freedom interacting on a lattice. They are also called {\em intrinsically localized}, in distinction, e.\,g., to Anderson localization triggered by disorder. A necessary condition for their existence is the nonlinearity of the equations of motion of the system, and the existence of discrete breathers has been proved rigorously for some classes of systems \cite{MKAub,Bambusi,LiSpiMK,AuKoKa,James,james2,JaNo,jamesnoble}. In contrast to their analogues in continuous systems, the existence of breathers in discrete systems is a generic phenomenon, which accounts for considerable interest in these objects from a physical point of view in the last decade. In fact, recent experiments could demonstrate the existence of discrete breathers in various real systems such as low-dimensional crystals \cite{Swanson_ea}, antiferromagnetic materials \cite{SchwarzEnSie}, Josephson junction arrays \cite{BiUs}, molecular crystals \cite{EdHamm}, micromechanical cantilever arrays \cite{Sato_ea}, and coupled optical waveguides \cite{Mandelik_ea}. The presence of discrete breathers has a significant influence on the properties of a system, and their relevance is discussed for a variety of physical effects, ranging from anomalous heat conduction to targeted energy transfer. For a review of some properties of discrete breathers see \cite{FlaWi}.

In the present paper, we study discrete breathers in the diatomic Fermi-Pasta-Ulam (FPU) chain, a chain of alternating light masses $m_1$ and heavy masses $m_2$, coupled by anharmonic forces. This model has an acoustic and an optic phonon band, separated by a gap. For hard interaction potentials, discrete breathers are known to exist either as ``optic breathers'' with frequencies above the optic band, or as ``acoustic breathers'' with frequencies in the gap between the acoustic and the optic band \cite{LiSpiMK,jamesnoble} (existence results are also available for soft potentials \cite{jamesnoble}). In general, for some given parameter values (breather frequency, mass ratio, parameters of the interaction potential), a variety of different breather solutions can exist, and there have been attempts towards a classification of discrete breathers in certain models \cite{AlfBraKo}. Here, we will study breathers of different types and bifurcations between them in the diatomic FPU chain, with breather frequency $\omega$ and mass ratio $m=m_1/m_2$ as bifurcation parameters. We will use both numerical and analytical methods to identify existence regions of different types of discrete breathers and their bifurcation lines in the small amplitude limit.

In particular we numerically identify a period tripling bifurcation around optic breathers, which leads to new breather solutions with frequencies in the gap, and a second local bifurcation around acoustic breathers. These results provide new breather solutions of the FPU system which interpolate between the classical acoustic and optic modes.

These numerical results are complemented by a local analytical study of the two bifurcations. One difficulty is that the two basic states consist in breather solutions, instead of having the FPU chain at rest as a starting state. Our approach uses the fact that the two bifurcation lines originate from a particular ``corner'' in the parameter space $\left(\omega^2,m\right)$. As parameters lie near this corner, we prove by means of a center manifold reduction that small amplitude solutions can be described by a four-dimensional reversible map. This allows us to derive formally a continuum limit differential equation which describes the numerically observed bifurcations at leading order. With this leading order analysis, the problem reduces to studying homoclinic bifurcations around reversible homoclinics of the asymptotic differential system. This allows one to compute the local shape of the ``sector'' in parameter space where breather bifurcations are numerically observed. In addition we provide leading order approximations of the new bifurcating breather solutions.
 
After introducing the diatomic FPU chain in section \ref{sec:FPU}, an overview regarding some existence results for discrete breathers in this model is given in section \ref{sec:DBFPU}. The first part of the paper is devoted to a numerical investigation, and we start by briefly reviewing the technique of numerical continuation of discrete breathers in section \ref{sec:numcont}. Numerical results obtained with this technique are presented in section \ref{sec:numresults}. The second part of the paper contains the complementary analysis of small amplitude discrete breathers by means of analytic methods. To this purpose, the equations of motion of the diatomic FPU chain are reformulated as a map in some loop space in section \ref{sec:loopspace}, and a center manifold reduction, allowing to locally reduce the dimensionality of the map, is performed in section \ref{sec:reduction}. Section \ref{sec:asymptotics} provides a formal asymptotic analysis for this reduced map. In section \ref{sec:comparison} this analytic result is confronted with the numerical findings, and we conclude with a summary and some comments in section \ref{sec:summary}.

\section{Diatomic FPU chain}\label{sec:FPU}

In 1955, E.~Fermi, J.~Pasta, and S.~Ulam published a study on the failure of short-time energy equipartition observed for certain initial conditions in a one-dimensional Hamiltonian lattice with nonlinear nearest-neighbor interactions \cite{FePaU}. Despite its simple form, this system reveals a surprisingly rich and complex behavior, and it has become one of the most studied models in the theory of nonlinear phenomena. Here we consider a diatomic extension of this model, consisting of a chain of alternate masses, coupled by identical nonlinear springs. The Hamiltonian function of the system is
\begin{equation}
\label{eq:FPU_Ham}
H = \sum_{n \in \Z} \left[ \frac{p^2_n}{2\,m_n} + W\left(q_{n+1}-q_n\right) \right],
\end{equation} 
with momenta $p_n\in\R$ and mass displacements $q_n\in\R$, and the interaction potential $W$ is of the form
\begin{equation}\label{eq:W}
W\left(x\right) = \frac{\omega_0^2}{2} x^2 + \frac{\beta}{4} x^4 +\Or \left( x^6 \right) ,
\end{equation}
where $\omega_0,\beta\in\R$ and $\Or $ is the order symbol. We assume $W$ to be an even function to simplify the computations, but all the analysis which follows could be extended to non even potentials containing a cubic term in the expansion \eref{eq:W}. Alternate masses correspond to the choices $m_{2n+1}=m_1$ and $m_{2n}=m_2$ for all $n\in\Z$, where, without loss of generality, we choose $m_1<m_2$. Whenever resorting to numerical methods, we fix $W\left(x\right) = \frac{1}{2} x^2 + \frac{1}{4} x^4$, $m_1=1$, $m_2 = m^{-1}$ (where $m \in (0,1)$ is a parameter) and refer to a finite FPU chain with free boundary conditions, i.\,e., a Hamiltonian function
\begin{equation}
\label{eq:FPU_Ham_finite}
H = \sum_{n=1}^N \frac{p^2_n}{2\,m_n} + \sum_{n=1}^{N-1} W\left(q_{n+1}-q_n\right)
\end{equation} 
consisting of $N$ degrees of freedom.

\section{Discrete breathers in the diatomic FPU chain}
\label{sec:DBFPU}

The first numerical observations of discrete breathers in diatomic FPU systems date back to the beginning of the 1990s \cite{BuKiPy,Aoki}. A first rigorous proof of the existence of discrete breathers in this system was obtained by Livi, Spicci, and MacKay \cite{LiSpiMK} in 1997. These authors showed for hard interaction potentials the existence of discrete breathers with frequencies above the optic band by continuation of a single-site breather, trivially existing in the ``uncoupled'' limit $m=m_1/m_2\to0$, towards small but non-zero mass ratios $m$. It was not until recently that the existence of small amplitude discrete breathers was proved for arbitrary mass ratios by James and Noble \cite{JaNo,jamesnoble} by means of a center manifold reduction. In the present article, this type of analysis is extended to the case of a higher dimensional center manifold. This extension allows to describe bifurcations between different types of discrete breathers occurring in certain parameter ranges.

\begin{figure}[hb]
\center
\includegraphics[width=10cm,keepaspectratio=true]{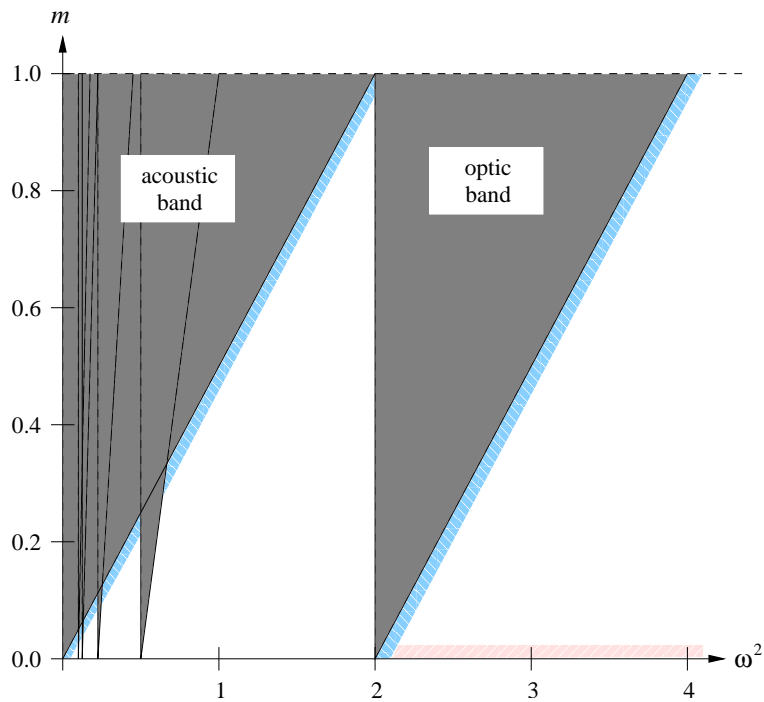}
\caption{\label{bands}Bands (shaded) in parameter space $\left(\omega^2,m\right)$ in which existence of discrete breathers is not expected due to resonance effects. The figure refers to the rescaled system \eref{fdi}, or equivalently to the original system \eref{eq:FPU_Ham} with $m_1 =1$, $m_2 = m^{-1}$ and $\omega_0=1$. The broad bands are the acoustic, respectively the optic, band of the linearized system. The narrower ones are images of the optic band, corresponding to breather frequencies $\omega$ whose multiples are resonant with the optic band. A countably infinite number of these images of the optic band exist. The images of the acoustic band are not plotted as they all lie inside the acoustic band and do not give any additional forbidden zones. Hatched areas sketch parameter values for which existence of discrete breathers has been proved for a hard interaction potential ('normally' hatched: proved in \cite{JaNo,jamesnoble}, lightly hatched: proved in \cite{LiSpiMK}).}
\end{figure}

At least formally, a necessary condition for discrete breathers to exist appears to be the nonresonance of the discrete breather frequency $\omega$ with the phonon spectrum $\omega_q$ of the linearized system, i.\,e., $\omega_q/\omega\notin\mathbbm{N}$ \cite{Flach}. The plot of figure \ref{bands} serves to illustrate where forbidden zones in parameter space $\left(\omega^2,m\right)$, i.\,e., zones of frequencies $\omega$ not complying with this condition, lie.
James and Noble proved the existence of discrete breathers outside (but very close to) the bands as introduced in the caption of figure \ref{bands}. For so-called hard potentials with $\beta>0$, discrete breathers are found slightly above the bands, in the case of soft potentials with $\beta<0$ slightly below. For hard potentials the zones of proved breather existence are sketched in figure \ref{bands} (hatched areas). The result of Livi, Spicci, and MacKay \cite{LiSpiMK}, in contrast, shows existence of discrete breathers for arbitrary $\omega^2>2$, but close to the base line at $m\gtrsim 0$ (lightly hatched in figure \ref{bands}). Although rigorous existence proofs apply only to narrow regions in parameter space, numerical results indicate that discrete breathers can be found for arbitrary parameter values outside the forbidden zones \cite{CreLiSpi}.

\section{Numerical continuation of discrete breathers}
\label{sec:numcont}

The focus of our interest lies on the various types of symmetric discrete breathers, their regions of existence in parameter space, and on the bifurcations occurring between them. In the first half of this article, with the intention to attain an overview of the types of discrete breathers that can be expected and on where bifurcations might show up, we resort to numerical techniques. The technique we use to compute discrete breathers up to numerical accuracy was put forward and discussed by Mar\'{\i}n and Aubry in \cite{MaAub}. It is based on Newton's method to compute zeros of functions from a starting value sufficiently close to the desired zero. In this way, families of discrete breathers can be constructed by ``continuing'' a known breather solution of certain parameter value $\lambda$ [a vector, in our case $(\omega^2,m)$, of any continuously varying parameters characterizing the solution] to a close-by value $\lambda'$. This continuation procedure has been put on rigorous grounds by Sepulchre and MacKay \cite{MacSep}, by introducing the notion of ``normal'' periodic orbits which can be shown to persist for small continuous parameter changes.

\subsection{Seeds for numerical continuation}
With this continuation method at our disposal, and in order to compute a certain family of discrete breathers, this leaves us with the problem of finding a starting value (or ``seed'') for the numerical continuation, appropriate for the breather family of interest. The standard procedure to obtain such a seed is the anti-continuous limit, as mentioned in section \ref{sec:DBFPU} and utilized for the first existence proof \cite{LiSpiMK} of discrete breathers in the diatomic FPU chain. In the limit $m_2\to\infty$ when every second mass is infinite, time periodic and spatially localized solutions exist trivially. The simplest one consists in the oscillation of one particle between ``fixed walls'' (two infinite masses), while all the other particles are at rest. However, for hard potentials \eref{eq:W} with $\beta>0$, the continuation of such a seed will result in discrete breathers with frequencies above the optic band. This happens due to the fact that, in figure \ref{bands}, such a seed corresponds to a point on the line $\left(\omega^2,m\right)=\left(\omega^2,0\right)$ with $\omega^2>2$, and, as discrete breathers cannot be continued within the bands, a continuation from this seeds to frequencies with $\omega^2<2$ is in general not possible.

The bifurcations between different types of discrete breathers we are interested in will be found prevalently in the gap in parameter space $\left(\omega^2,m\right)$ between the acoustic and the optic band. The seeds necessary to compute gap breathers by the above described numerical continuation method can be obtained by several strategies.

\begin{figure}
\center
\includegraphics[width=8cm,clip=true,keepaspectratio=true]{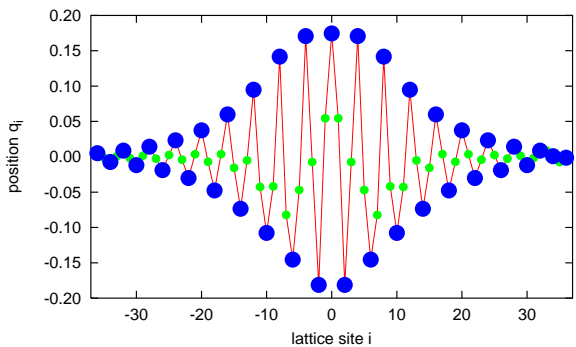}
\caption{\label{RWA} Snapshot of a gap breather at the instant of maximum oscillation amplitude for a diatomic FPU chain with coupling constants $\omega_0 =1$, $\beta=1$ and $\left(\omega^2,m\right)\approx\left(0.256,0.126\right)$, computed numerically with a seed from the rotating wave approximation. The large masses $m_2$ (large dots) oscillate with larger amplitude than the small masses $m_1$ (small dots).
}
\end{figure}

\begin{figure}[hb]
\center
\includegraphics[width=11cm,clip=true,keepaspectratio=true]{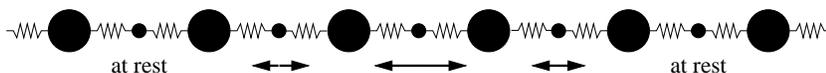}
\caption{\label{fig:commensillu} Illustration of a seed for a gap breather from the anti-continuous limit with commensurate frequencies. The large balls indicate large masses $m_2\to\infty$, coupled to the small masses (smaller balls) by nonlinear springs. All masses except the three light ones indicated are at rest. The middle mass oscillates with larger amplitude, and therefore smaller period, than the two adjacent oscillating small masses.}
\end{figure}

\begin{figure}[hb]
\center
\includegraphics[width=8cm,clip=true,keepaspectratio=true]{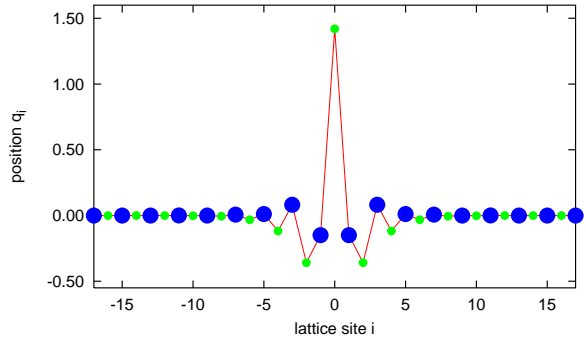}
\caption{\label{fig:commensshape} Snapshot of a gap breather at the instant of maximum oscillation amplitude for a diatomic FPU chain with coupling constants $\omega_0 =1$, $\beta=1$ and $\left(\omega^2,m\right)\approx\left(0.253,0.689\right)$, obtained by using a seed from the anti-continuous limit with commensurate frequencies $2\omega$ and $3\omega$. The large masses $m_2$ (large dots) oscillate with smaller amplitude than the small masses $m_1$ (small dots).
}
\end{figure}

One way is to use a rotating wave approximation which consists in approximating discrete breathers by using a small number of Fourier modes \cite{SieTa,franc1,franc2}. In the small amplitude limit, the slow spatial modulation of the fundamental harmonic is approximately determined by a nonlinear Schr\"odinger equation \cite{Konotop,nikos}. This provides an explicit approximation of the envelope in the form of a soliton profile, which serves as an initial guess for the Newton method. For hard potentials, the computed starting states are of the ``acoustic breather type'', i.\,e., similarly to the acoustic modes of the linearized equations of motion, the large masses $m_2$ oscillate at large amplitude, while the small ones $m_1$ move only slightly (see figure \ref{RWA}).
An entirely different strategy to obtain seeds for the continuation of gap breathers makes use of the anti-continuous limit with big masses $m_2$ set to infinity in a slightly more sophisticated way than explained in section \ref{sec:numcont}. Instead of the simplest case with only one oscillating light mass $m_1$, two or more oscillating ones with different frequencies are considered. Taking, for example, a light mass oscillating with period $T_1$, while its two adjacent light masses oscillate with $T_2=\frac{3}{2}T_1$, a discrete breather with period $T=3T_1$ is obtained (see figure \ref{fig:commensillu} for an illustration).
In this way, although $T_1,T_2<\pi\sqrt{2}$, the overall period $T$ of the discrete breather can be larger than $\pi\sqrt{2}$, giving rise to a gap breather. An example of a discrete breather produced by continuation of such a seed to finite mass ratios is provided in figure \ref{fig:commensshape}. In a similar way, more than three oscillating particles and other frequency ratios can be employed to obtain a large variety of gap breathers.

\subsection{Boundary conditions and chain length}
As specified in section \ref{sec:FPU}, free boundary conditions have been used for the numerical computation of discrete breathers. Under reasonable conditions on the chain length, i.\,e., on the number $N$ of masses, this choice of boundary conditions does not affect the results reported: When the chain length is sufficiently large with respect to the localization length of the discrete breather, the oscillation amplitudes close to the boundaries are anyway smaller than the precision (typically of order $10^{-16}$) of the numerical computation. As a consequence, the chain length necessary for the accurate numerical computation of a discrete breather increases with its localization length, and it is this need for large-$N$-chains which imposes a limitation on the parameter values $(\omega^2,m)$ for which the numerical computation of small amplitude discrete breathers is possible. All numerical results plotted in this article were obtained for chains consisting of $N=99$ masses.

\section{Numerical results}
\label{sec:numresults}

With the above described numerical methods at hand, bifurcations between various types of discrete breathers are observed and investigated: In section \ref{sec:symasymbifu}, the bifurcation line of a bifurcation between symmetric and asymmetric discrete breathers is determined. In section \ref{sec:cornerbifu}, discrete breathers are investigated for parameter values $\left(\omega^2,m\right)$ close to a corner formed by the acoustic band and an image of the optic band (see figure \ref{bands}). Two kinds of bifurcations are detected in this region of parameter space: a period-multiplying bifurcation on the one hand, and a local bifurcation around breathers whose oscillation is mainly performed by the large masses $m_2$ on the other hand. Each of these bifurcations is of pitchfork type, and the bifurcation lines in the $\left(\omega^2,m\right)$-plane are computed in each case. All numerical results reported in this section are for coupling constants $\omega_0 =1$, $\beta=1$ in the interaction potential \eref{eq:W}.

\subsection{Symmetric--asymmetric bifurcation}
\label{sec:symasymbifu}

\begin{figure}[ht]
\center\hspace{-1mm}
\includegraphics[width=5.5cm,clip=true,keepaspectratio=true]{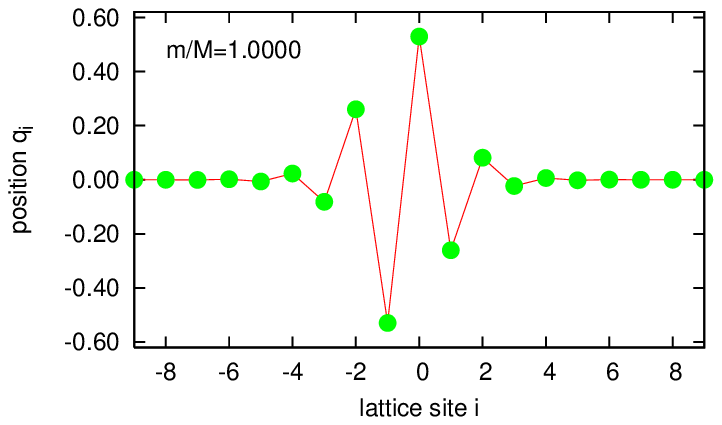}\hspace{-5mm}
\includegraphics[width=5.5cm,clip=true,keepaspectratio=true]{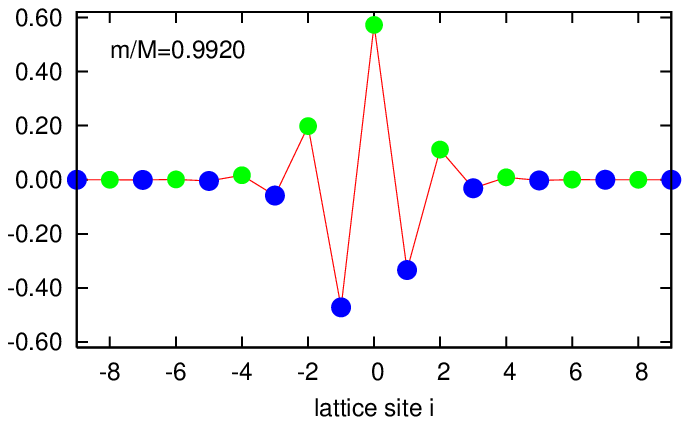}\hspace{-5mm}
\includegraphics[width=5.5cm,clip=true,keepaspectratio=true]{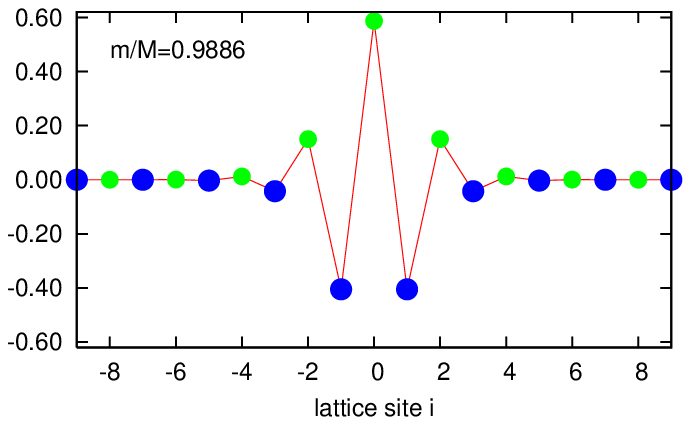}\hspace{-1mm}
\caption{\label{fig:DBshapesymasym} Snapshots of breathers from the same breather family at the instant of maximum oscillation amplitude for a diatomic FPU chain with coupling constants $\omega_0 =1$, $\beta=1$ for parameter values $\omega^2=5.8397$ and $m\in\{1.0000,0.9920,0.9886\}$. The linearly stable antisymmetric, bond-centered breather (Page mode, left figure) at mass ratio $m=1$ deforms into an asymmetric breather for $m\lesssim1$ (middle figure). For sufficiently small mass ratios, this branch of asymmetric breathers and the branch of symmetric, site-centered breathers (Sievers-Takeno mode, right figure) collide and a bifurcation takes place.}
\end{figure}

At the end of the 1990s, a bifurcation between symmetric and asymmetric discrete breathers in the diatomic FPU chain was reported by Cretegny, Livi, and Spicci \cite{CreLiSpi}. These authors investigated two relatively simple types of discrete breathers, the so-called Sievers-Takeno mode and the Page mode (see figure \ref{fig:DBshapesymasym} for an illustration of these modes), above the optic band.
%
They performed a linear stability analysis of these modes, observing that for small values of the mass ratio $m$ the Sievers-Takeno mode is linearly stable, while for $m=1$ the Sievers-Takeno mode is unstable and the Page mode is linearly stable (see \cite{CreLiSpi} for details of the stability analysis). This observation led to the conjecture, and the subsequent numerical confirmation, of the existence of a bifurcation connecting these two branches of solutions. We give more informations about this bifurcation by computing numerically the bifurcation line in the $\left(\omega^2,m\right)$-plane.

Since the bifurcation takes place above the optic band, the anticontinuous limit can be used as a seed for the numerical continuation. Discrete breathers of the Sievers-Takeno-type are obtained from the ``standard'' anticontinuous limit of alternate finite masses $m_1$ and infinite masses $m_2$, where only one of the light masses $m_1$ oscillates while all the others are at rest. Numerically, the continuation is possible across the entire range of mass ratios $0\leqslant m\leqslant1$.

The computation of the Page mode in the diatomic chain is slightly more complex using the anticontinuous limit. This requires a preliminary step, where discrete breathers of the Page-type are produced from an anticontinuous limit where every third mass is infinite while all other masses are equal and finite. Continuing a seed, with two adjacent light masses oscillating out of phase and all others at rest, up to mass ratio $m=1$ (i.\,e., all masses in the chain are equal), the Page mode is obtained. Then, starting from mass ratio one, the Page mode can be continued to the diatomic chain of alternate masses. 

\begin{figure}
\psfrag{m}{{\small $m$}}
\psfrag{w}{{\small $\omega^2$}}
\psfrag{2}{}
\center
\includegraphics[width=9cm,height=6cm,clip=false,keepaspectratio=true]{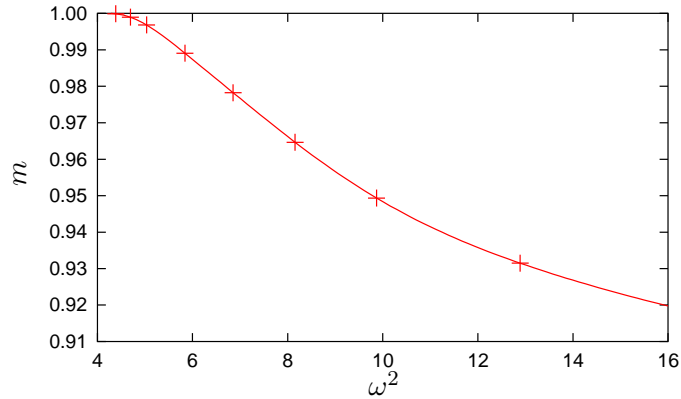}
\caption{\label{fig:bifupoints} Bifurcation line in the parameter plane $\left(\omega^2,m\right)$ for the symmetric--asymmetric bifurcation of discrete breathers. Below the bifurcation line, linearly stable symmetric site-centered breathers (Sievers-Takeno mode) are observed. At the bifurcation line these breathers become unstable and, additionally, linearly stable asymmetric breathers show up. For equal masses $m=1$ these asymmetric ones transform into anti-symmetric bond-centered breathers (Page mode). Lines connecting the numerically determined data points are drawn to guide the eye.}
\end{figure}

In figure \ref{fig:bifupoints}, the bifurcation line for the bifurcation between symmetric and asymmetric discrete breathers is plotted in a section of the $\left(\omega^2,m\right)$-plane.
Below the bifurcation line, discrete breathers of the Sievers-Takeno type are observed and are found to be linearly stable. At the bifurcation line, these breathers loose linear stability and, as additional solutions, asymmetric breathers emerge. Above the bifurcation line, unstable Sievers-Takeno type breathers and linearly stable asymmetric breathers coexist. At mass ratio $m=1$, the asymmetric breathers transform into antisymmetric breathers of the Page type. Note that, in the diagram sketching the phonon bands of the linearized equations of motion (figure \ref{bands}), the bifurcation line emanates from the endpoint of the upper band edge of the optic band. It is a common feature of all the bifurcation lines investigated in this article that they emanate from endpoints of bands or corners formed by bands and their images as shown in figure \ref{bands}. 

\subsection{Bifurcations close to ``corners'' formed by bands and their images}
\label{sec:cornerbifu}

In reference \cite{JaNo,jamesnoble}, James and Noble proved the existence of discrete breathers for the diatomic FPU chain in the vicinity of edges of the phonon bands, with different functional forms of the discrete breathers close to the different band edges. At the ``corners'' in parameter space $\left(\omega^2,m\right)$ formed by the acoustic band and the images of the optic band (see figure \ref{bands}), one might conjecture (for hard potentials) the occurrence of bifurcations connecting these two types of breathers, i.\,e., an optic and an acoustic breather. This conjecture was the original motivation for the study of gap breathers reported in this article.

\begin{figure}[ht]
\center
\begin{minipage}{10.4cm}\center
\includegraphics[width=5.1cm,clip=false,keepaspectratio=true]{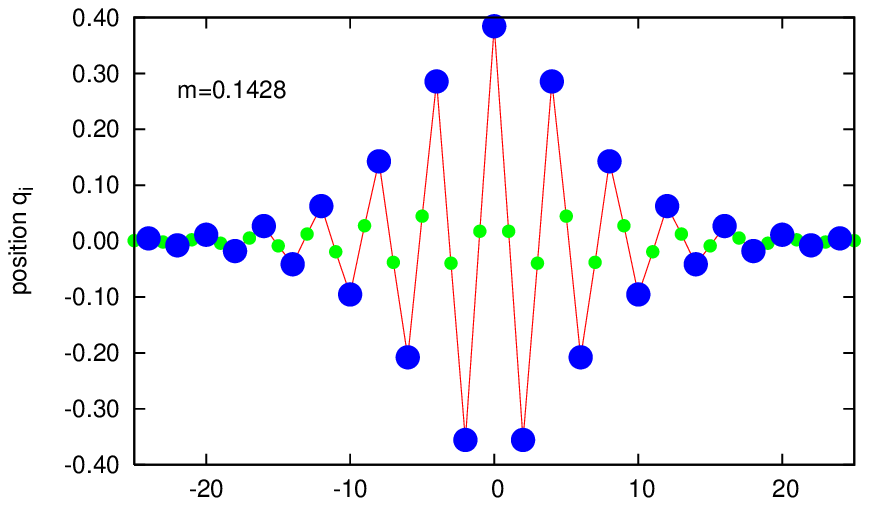}\\
\includegraphics[width=5.1cm,clip=false,keepaspectratio=true]{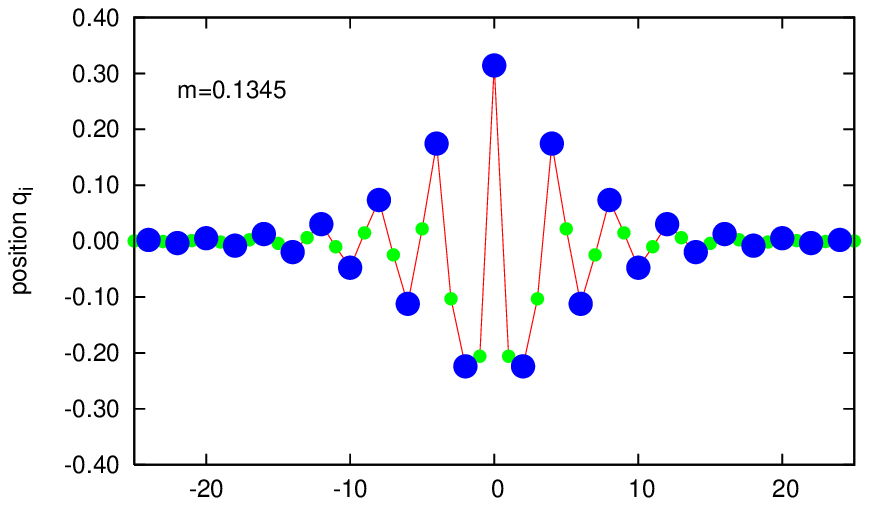}\hspace{-1mm}
\includegraphics[width=5.1cm,clip=false,keepaspectratio=true]{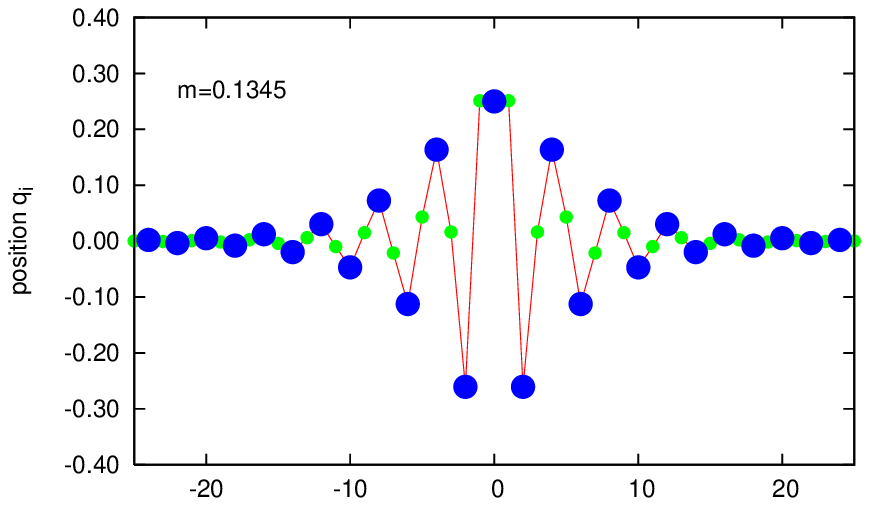}\\
\includegraphics[width=5.1cm,clip=false,keepaspectratio=true]{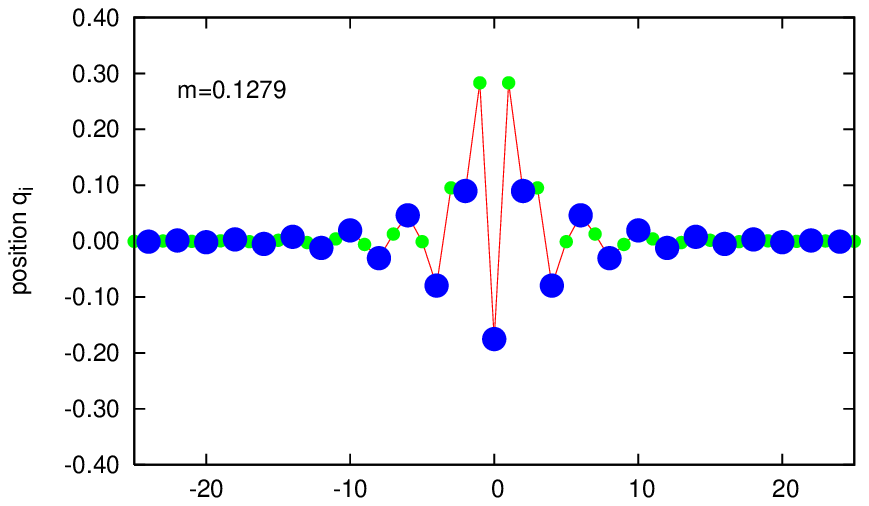}\hspace{-1mm}
\includegraphics[width=5.1cm,clip=false,keepaspectratio=true]{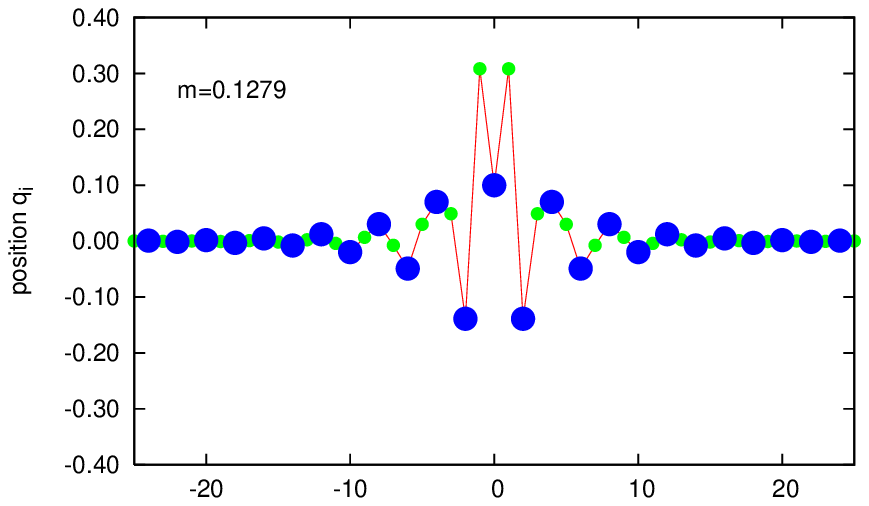}\\
\includegraphics[width=5.1cm,clip=false,keepaspectratio=true]{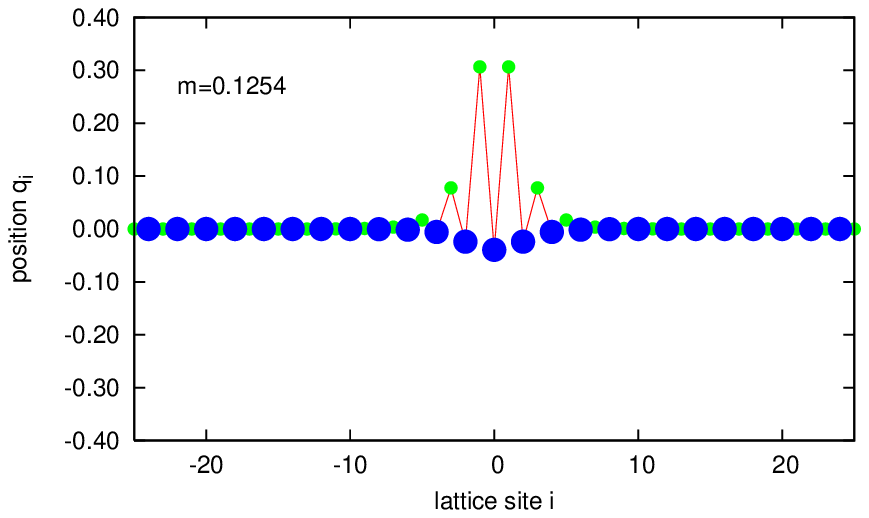}
\end{minipage}
\hfill
\begin{minipage}{5.1cm}\center
\includegraphics[width=5.1cm,clip=false,keepaspectratio=true]{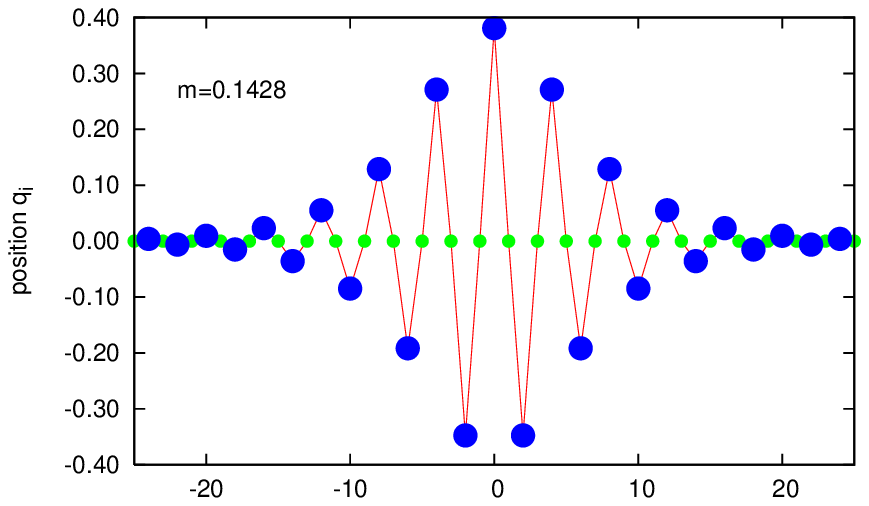}
\includegraphics[width=5.1cm,clip=false,keepaspectratio=true]{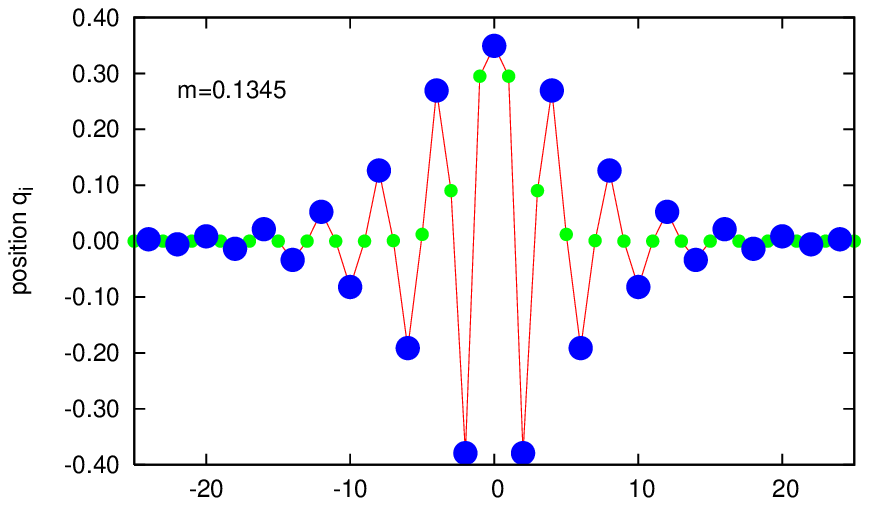}
\includegraphics[width=5.1cm,clip=false,keepaspectratio=true]{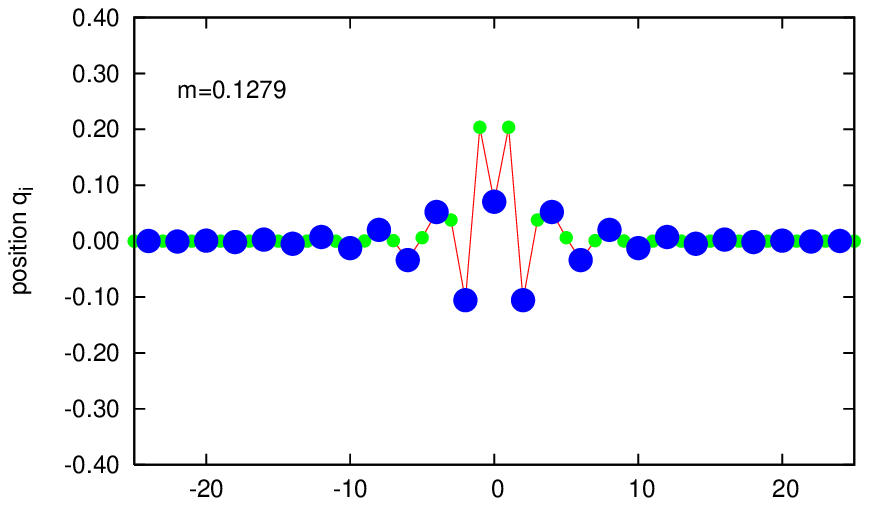}
\includegraphics[width=5.1cm,clip=false,keepaspectratio=true]{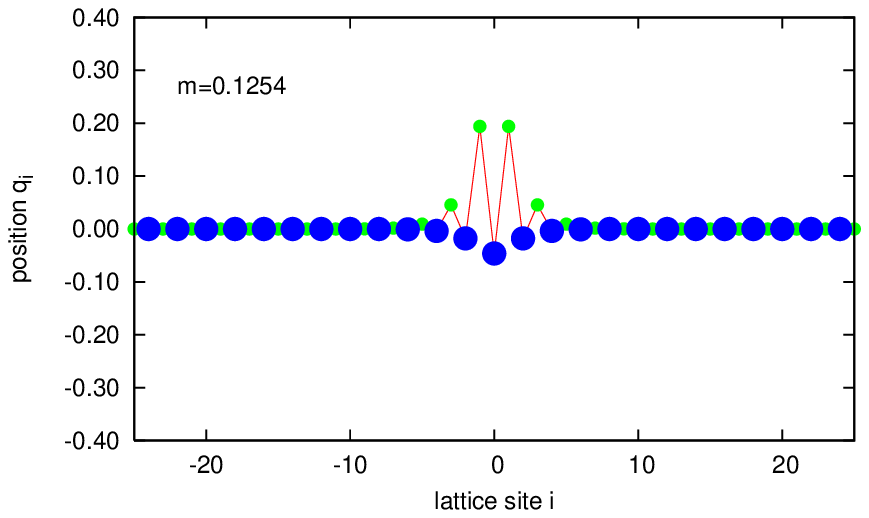}
\end{minipage}
\caption{\label{fig:RWAshapeM<=>m} Snapshots of gap breathers from the same breather family at the instant of maximum oscillation amplitude for a diatomic FPU chain with coupling constants $\omega_0 =1$, $\beta=1$ for different points on the line $\left(\omega^2,m\right)=\left(2m+0.01555,m\right)$ in parameter space. For $m\approx 0.1428$, the large masses $m_2$ (large dots) have a larger oscillation amplitude, while for decreasing $m$ the amplitude of the small masses' oscillations (small dots) increases. The left and middle columns are the results of a numerical computation of discrete breathers from two different bifurcation branches. The right column shows approximate analytic results as obtained for the indicated parameter values from equations \eref{acousticpair}, \eref{acousticpairbif}, \eref{opticpairbif}, and \eref{opticpair} (from top to bottom) for comparison with the numerical results in the middle column (see section \ref{sec:comparison} for details). The qualitative agreement of numerical and analytic results is excellent throughout. The quantitative agreement, as expected, is good only for weakly localized breathers (e.\,g.\ in the upper row) for which the assumptions of the analytic calculation are met. For parameter values $\left(\omega^2,m\right)$ even closer to the phonon band edges for which also the breather types shown in the lower row would be less localized, we encountered problems with the convergence of the algorithm used for the numerical computation and therefore could not compare analytic and numerical results.}
\end{figure}

We chose the corner formed by the acoustic band and the ``second image'' of the optic band at $\left(\omega^2,m\right)=\left(\frac{1}{4},\frac{1}{8}\right)$ (see figure \ref{bands}) for an exemplary investigation. For the numerical detection of bifurcations in this region we utilized seeds obtained from the rotating wave approximation for the numerical continuation. Two types of local bifurcations of pitchfork type were found:
\begin{enumerate}
\item a period multiplying bifurcation (local bifurcation around an optic breather whose oscillation is mainly performed by the small masses $m_1$),
\item a local bifurcation around an acoustic breather whose oscillation is mainly performed by the large masses $m_2$.
\end{enumerate}
As parameters are varied, these two bifurcations connect discrete breathers with prevalent oscillations of the heavy masses $m_2$ to discrete breathers with prevalent oscillations of the light masses $m_1$. In the left part of figure \ref{fig:RWAshapeM<=>m}, the transformations of the shape of discrete breathers successively undergoing the two bifurcations are illustrated for two different bifurcation branches. The right column of figure \ref{fig:RWAshapeM<=>m} compares these numerical solutions to analytical approximations obtained in sections \ref{sec:asymptotics} and \ref{sec:comparison} (these results will be commented in section \ref{sec:comparison}).

\begin{figure}[ht]
\psfrag{m}{{\small $m$}}
\psfrag{g}{{\small $g$}}
\center
\includegraphics[width=9cm,clip=false,keepaspectratio=true]{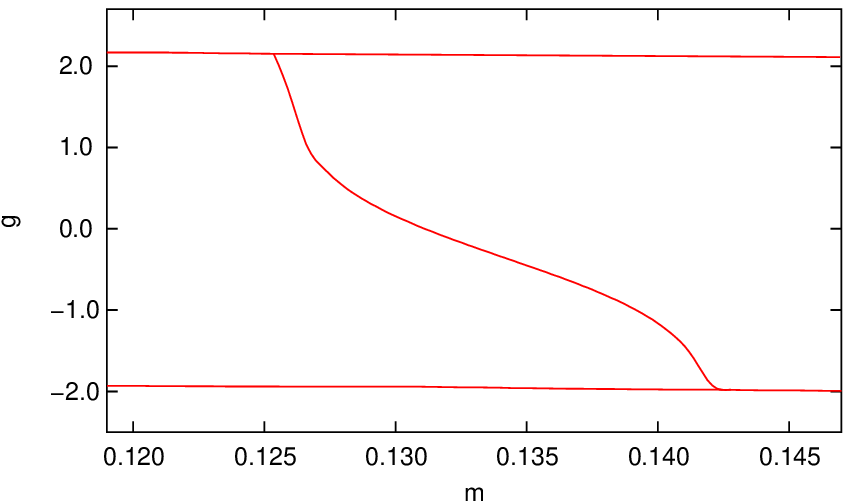}
\caption{\label{fig:doublebifu} 
Along the line $\left(\omega^2,m\right)=\left(2m+0.01555,m\right)$ in parameter space, the quantity $g$ defined in \eref{eq:g} is plotted for discrete breathers from different bifurcation branches as a measure of how big the contributions of the light and the heavy masses, respectively, are to the overall oscillation. The numerical continuation of solutions connects discrete breathers with predominant heavy and light mass oscillations (top and bottom lines) via a succession of two pitchfork bifurcations. Computations are performed for a diatomic FPU chain with $N=99$ degrees of freedom and coupling constants $\omega_0 =1$ and $\beta=1$.
}
\end{figure}

As a measure of how strongly the oscillation of the discrete breather is dominated by the heavy or the light masses, the quantity
\begin{equation}\label{eq:g}
g=\frac{1}{2}\ln\frac{\sup\limits_t \sum\limits_n q_{2n\vphantom{+}}^2\left(t\right)}{\sup\limits_t\sum\limits_n q_{2n+1}^2\left(t\right)}
\end{equation}
may be used. We consider here the ratio $r$ of the mean-square displacement of heavy masses by the mean-square displacement of the light ones, whose logarithm is given by $g = \ln r$. A positive value of $g$ corresponds to a discrete breather dominated by the oscillation of the large masses, whereas a negative value of $g$ indicates that the oscillation of the small masses is dominating. Plotting $g$ for a family of discrete breathers (i.\,e., as a function of the parameter labeling the members of the family), the interpolation between acoustic and optic breathers provided by the bifurcating solutions can be nicely illustrated (see figure \ref{fig:doublebifu}). We have plotted $g$ instead of $r-1$ in order to stretch the region where $r<1$, thereby increasing the visibility of this region in the plot.

\begin{figure}[hb]
\psfrag{w}{{\small $\omega^2$}}
\psfrag{b}{}
\psfrag{2}{}
\psfrag{m/M}{{\small $m$}}
\center
\includegraphics[width=9cm,clip=false,keepaspectratio=true]{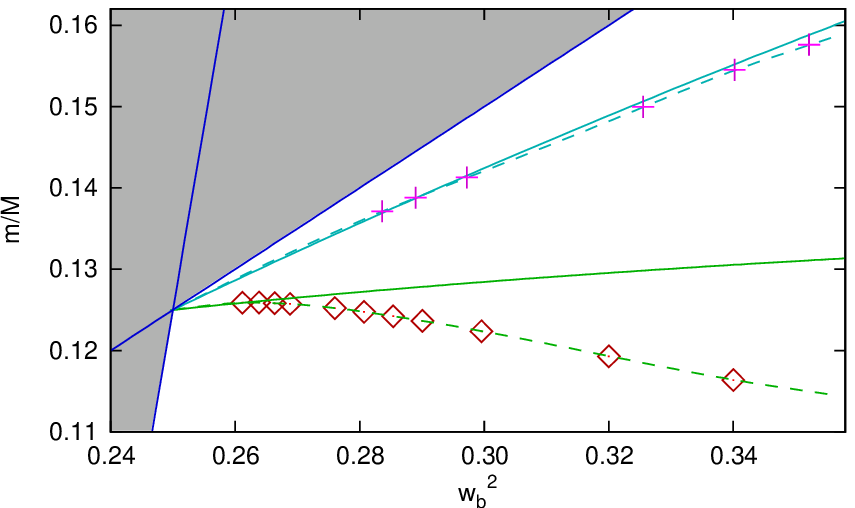}
\caption{\label{fig:bifudiagram} Bifurcation lines close to the corner at $\left(\omega^2,m\right)=\left(\frac{1}{4},\frac{1}{8}\right)$ for the period tripling bifurcation (lower lines and $\Diamond$) and the local bifurcation around an acoustic breather (upper lines and $+$). Above the upper bifurcation line, discrete breathers with predominant heavy mass oscillations are observed. At the bifurcation line, additional breathers show up with light mass oscillations growing in importance. At the lower bifurcation line, the frequencies of the breathers with light mass oscillation are tripled which, in the plotted parameter plane, corresponds to a jump to frequencies above the optic band (see figure \ref{bands}). Dashed lines connecting the numerically determined data points are drawn to guide the eye. The solid lines are the analytic approximations \eref{critpt} (lower line) and \eref{critptbis} (upper line) of the bifurcation lines obtained by means of a center manifold reduction, valid in the vicinity of the corner.
}
\end{figure}

For both, the local bifurcation around an acoustic breather and the period tripling bifurcation, the bifurcation lines in the parameter plane $\left(\omega^2,m\right)$ have been computed numerically and are plotted in figure \ref{fig:bifudiagram}.
Reliable data points even closer to the bands could not be obtained numerically, as the localization length of the discrete breather is increasing when approaching the band and therefore size effects in the finite systems used for the numerics become non-negligible.

\section{Center manifold reduction near critical parameter values}

The bifurcations observed numerically in the previous section were found to lie in the vicinity of crossings of phonon band edges, supporting the conjecture that bifurcation lines emanate from the corners formed by these bands. In this section we apply the technique of center manifold reduction for an analytic investigation of these bifurcations. When proving the existence of discrete breathers close to the phonon band edges in the diatomic FPU chain in \cite{JaNo,jamesnoble}, James and Noble rewrote the equations of motion of the diatomic FPU chain as a map in some loop space and accomplished a reduction of the spatial dynamics onto a two-dimensional center manifold. In the vicinity of the ``corners'' in parameter space we are interested in, a similar analysis can be performed, with the complication that the relevant center manifold will be four dimensional. In section \ref{sec:asymptotics} we shall analyze asymptotically the four-dimensional map on the center manifold. A set of equations will be obtained whose solutions correspond to different types of discrete breathers, some of which can be identified as the ones observed numerically in the previous section.


\subsection{Formulation of the FPU system as a map in a loop space}
\label{sec:loopspace}

We consider the diatomic FPU chain \eref{eq:FPU_Ham} consisting of an infinite number of degrees of freedom, and reformulate this system as a map in loop space. We start from the equations of motion 
\begin{equation}\label{fdintro}
m_{n}\,\frac{\rmd^2}{\rmd t^2}{q_{n}}=W'\left(q_{n+1}-q_{n}\right)-W'\left(q_{n}-q_{n-1}\right),\qquad n\in\Z,
\end{equation}
where $m_{2n+1}=m_{1}$, $m_{2n}=m_{2}$ ($m_1< m_2$), and the $q_{n}$ represent mass displacements from their reference position. $W$ is the interaction potential as defined in \eref{eq:W}, and a prime denotes its derivative. In order to reduce the number of parameters, we rescale \eref{fdintro} by setting $q_{n}\left(t\right)=\omega_0\,X_{n} \left(\omega_0\, t/\sqrt{m_{1}}\right)$, which yields
\begin{equation}\label{fdi}
\eqalign{
m^{-1}\frac{\rmd^2}{\rmd t^2}{X_{2n}}=V'\left(X_{2n+1}-X_{2n}\right)-V'\left(X_{2n}-X_{2n-1}\right),\\
\frac{\rmd^2}{\rmd t^2}{X_{2n+1}}=V'\left(X_{2n+2}-X_{2n+1}\right)-V'\left(X_{2n+1}-X_{2n}\right),
}
\end{equation}
with mass ratio $m=m_{1}/m_{2}\in \left(0,1\right)$ and
\begin{equation}\label{eq:V2}
V\left(x\right)=\frac{1}{2}x^2+\frac{\beta}{4}x^4+\Or \left( x^6 \right).
\end{equation}
Being interested in discrete breathers, we look for time-periodic solutions of \eref{fdi} with frequency $\omega$. In order to work in a space of functions with fixed period $2\pi$, we rescale $X_n$ by setting $X_n \left(t\right)=x_n \left(\omega t\right)$ ($x_n$ being $2\pi$-periodic), which transforms the equations of motion \eref{fdi} into
\begin{equation}
\label{newvar}
\eqalign{
m^{-1}\omega^2\frac{\rmd^2}{\rmd t^2}{x_{2n}}=V'\left(x_{2n+1}-x_{2n}\right)-V'\left(x_{2n}-x_{2n-1}\right),\\
\omega^2\frac{\rmd^2}{\rmd t^2}{x_{2n+1}}=V'\left(x_{2n+2}-x_{2n+1}\right)-V'\left(x_{2n+1}-x_{2n}\right).
}
\end{equation}
Breather solutions of \eref{fdi} with frequency $\omega$ correspond to $2\pi$-periodic solutions of \eref{newvar} satisfying the localization condition $\lim_{|n|\to\infty}{x_{n}\left(t\right)}=0$. For the center manifold reduction procedure, however, we do not need assumptions on the spatial behaviour of solutions, except the assumption of small amplitude.

Using the fact that $V^\prime$ is locally invertible, one can formulate \eref{newvar} as a first order recurrence relation in a loop space. Indeed, the first equation in \eref{newvar} defines a map $\left(x_{2n} ,x_{2n-1} \right)\mapsto x_{2n+1}$, the second one a map $\left(x_{2n} ,x_{2n+1} \right)\mapsto x_{2n+2}$. More explicitly, consider $u_{n}=x_{2n}$, $v_{n}=x_{2n-1}$ and $Y_{n}=\left(u_{n},v_{n}\right)$. Then, \eref{newvar} can be rewritten as
\begin{equation}
\label{recrel}
Y_{n+1}=F_{m,\omega}\left(Y_{n}\right) 
\end{equation} 
with
\begin{equation}
F_{m,\omega}\left(u,v\right)=\left(\!
    \begin{array}{c}
    G_{\omega^2}\left[ G_{\omega^2 /m }\left(u,v\right) , u \right] \\
    G_{\omega^2 /m }\left(u,v\right)
    \end{array}\!\right) 
\end{equation}
and 
\begin{equation}
G_{a}\left(u,v\right)={\left(V^\prime \right)}^{-1}\left[V^\prime \left(u-v\right)+a\, \frac{\rmd^2 u}{\rmd t^2}\right]+u.
\end{equation}
Note that the fixed point $\left(u,v\right)=0$ of $F_{m,\omega}$ corresponds to the FPU chain at rest. In the following, $m$ and $\omega$ play the role of bifurcation parameters. Even if the recurrence relation \eref{recrel} is infinite-dimensional and ill-posed (the operator $F_{m,\omega}$ is defined on a domain $\mathbbm{D}$ specified below), small amplitude solutions lie on a finite-dimensional center manifold whose dimension depends on the number of resonant phonons in the parameter regime considered \cite{james2,jamesnoble,JaNo}. The reduction to this finite-dimensional center manifold is the crucial step which allows for an analytic treatment of discrete breathers close to certain phonon band edges.

As a next step, we define appropriate function spaces on which $F_{m,\omega}$ is acting. We restrict our attention to even functions of $t$, thereby reducing the dimension of the center manifold by a factor of two (half of the Fourier modes are eliminated). Moreover, we look for $2\pi$-periodic oscillations consisting entirely of odd order harmonics $\cos\left[\left(2k+1\right)t\right]$ with $k\in\N_0$, which is possible due to the evenness of $V$. To be more precise, we introduce the function spaces $H^{n}_{\#}$ ($n\geqslant 0$), defined by
\begin{equation}
H^{n}_{\#}
=\left\{ u\in 
H^{n}\left(\mathbbm{R}/2\pi\mathbbm{Z}\right)\,\big|\,u\left(-t\right)=u\left(t\right),u\left(t+\pi \right)=-u\left(t\right)\right\},
\end{equation}
where $H^n$ denotes the standard Sobolev space. For $F_{m,\omega}\left(u,v\right)$ to be well-defined, we look for $\left(u_{n},v_{n}\right)\in\mathbbm{D}=H^{5}_{\#}\times H^{3}_{\#}$, and then the recurrence relation \eref{recrel} holds in $\mathbbm{X}=H^{1}_{\#}\times H^{3}_{\#}$. Due to the evenness of $V$ the operator $F_{m,\omega}$ maps $\mathbbm{D}$ onto $\mathbbm{X}$ (even order Fourier harmonics are not generated by applying $F_{m,\omega}$), and $F_{m,\omega}$ is analytic in a neighborhood of $\left(u,v\right)=0$ in $\mathbbm{D}$. 

For even potentials, working in a function space with only odd order Fourier harmonics allows to extend the existence results obtained by James and Noble for acoustic breather solutions (see \cite{jamesnoble}, theorem 1 iii)). This procedure suppresses half of the forbidden zones in figure \ref{bands}, namely the ones corresponding to even multiples of the optic band. In the general case of asymmetric potentials, where all Fourier harmonics are involved, the map $F_{m,\omega}$ linearized at $\left(u,v\right)=0$ possesses a pair of imaginary eigenvalues on the unit circle when parameters lie in these forbidden zones, which causes an obstruction for a breather existence result (see \cite{jamesnoble}). These additional eigenvalues (corresponding to even order Fourier harmonics) are eliminated with the present functional setting.

\subsection{Reduction of the dimension near a resonance of order three}
\label{sec:reduction}

As for the numerical investigation in section \ref{sec:cornerbifu}, we choose for a detailed analysis as an example the region of the crossing of the upper band edge of the acoustic band with the upper band edge of the ``second image'' of the optic band at $\left(\omega,m\right)=\left(\frac{1}{2},\frac{1}{8}\right)$ [or $\left(\omega^2,m\right)=\left(\frac{1}{4},\frac{1}{8}\right)$ in figure \ref{bands}]. In other words, we investigate the region close to a frequency resonance of order three between the optic phonon with wave number $q=0$ and the acoustic phonon with wave number $q=\pi$.

In order to study the local dynamics of \eref{recrel} around $Y_n=0$, we first need to determine the spectrum of the linearized map ${\mathrm D}F_{m,\omega}\left(0\right)$ close to the unit circle. This linear analysis has been already performed in reference \cite{jamesnoble} using a different (but completely equivalent) reformulation of the FPU system. In what follows, we briefly recall the useful spectral properties in our analysis and refer to reference \cite{jamesnoble} for more details. For all parameter values $(m,\omega )$, the spectrum of ${\mathrm D}F_{m,\omega}\left(0\right)$ on the unit circle consists in a finite number of eigenvalues which can be deduced from the usual dispersion relation for linear phonons (see below). Moreover, this {\em central } part of the spectrum is isolated from the {\em hyperbolic} part. Therefore ${\mathrm D}F_{m,\omega}\left(0\right)$ satisfies the property of spectral separation required to apply the center manifold theorem \cite{james2}.

${\mathrm D}F_{m,\omega}\left(0\right)$ admits a pair of simple eigenvalues $e^{\pm i q}$ on the unit circle (with an associated invariant subspace spanned by the vectors $\left(\cos\left( kt\right),0\right)$ and $\left(0,\cos\left( kt\right)\right)$, for an odd integer $k$) if $k \omega$ lies inside one of the acoustic or optic phonon bands plotted in figure \ref{bands}. With this pair of eigenvalues, equation \eref{fdi} linearized at $X_n=0$ admits solutions of the form $(X_{2n},X_{2n+1})=\cos\left( k \omega t\right) \, \left[ \, e^{iqn}\, \zeta +  e^{-iqn}\, \bar{\zeta} \, \right]$ for appropriate eigenvectors $\zeta \in \mathbb{C}^2$. These solutions correspond to linear standing waves with wave number $q$, where $q$ and $k\omega$ need to satisfy the classical dispersion relation for the diatomic FPU chain. This observation allows us to determine the structure of the spectrum in a rather simple way. For $\left(\omega,m\right)=\left(\frac{1}{2},\frac{1}{8}\right)$, the part of the spectrum of ${\mathrm D}F_{m,\omega}\left(0\right)$ which lies on the unit circle consists only of two double non-semi-simple eigenvalues $-1$ and $+1$, which split generically into two pairs of simple eigenvalues as parameters are slightly varied. The invariant subspace associated with eigenvalue $-1$ is spanned by the vectors $\left(\cos\left(t\right),0\right)$ and $\left(0,\cos\left(t\right)\right)$, whereas the invariant subspace belonging to eigenvalue $+1$ is spanned by $\left(\cos\left(3t\right),0\right)$ and $\left(0,\cos\left(3t\right)\right)$. If $\left(\omega,m\right)=\left(\frac{1}{2},\frac{1}{8}\right)$, $\omega$ lies at the top of the acoustic phonon band (corresponding to the wave number $q=\pi$), $3 \omega$ lies at the top of the optic band (wave number $q=0$), and no higher multiple of $\omega$ lie in the phonon spectrum. With this choice of parameters $\left(\omega ,m\right)$, the {\em central}\/ subspace (i.\,e., the subspace which is invariant under ${\mathrm D}F_{m,\omega}\left(0\right)$ and which is associated with the eigenvalues lying on the unit circle) is four-dimensional. 

In the following, we set 
\begin{equation}
\label{defparam}
m^{-1}\omega^2 = 2+\mu,\qquad
\omega^2 =\tfrac{1}{4}+\eta ,
\end{equation}
with $\mu,\eta \approx 0$, which implies $\omega\approx\frac{1}{2}$ and $m\approx \frac{1}{8}$. \Eref{recrel} [see also \eref{newvar}] can be rewritten as
\begin{equation}
\label{equv}
\eqalign{
\left(2+\mu \right)\, \frac{\rmd^2}{\rmd t^2}{u_{n}}=V'\left(v_{n+1}-u_{n}\right)-V'\left(u_{n}-v_{n}\right),\\
\left(\tfrac{1}{4}+\eta \right)\,\frac{\rmd^2}{\rmd t^2}{v_{n+1}}=V'\left(u_{n+1}-v_{n+1}\right)-V'\left(v_{n+1}-u_{n}\right).
}
\end{equation}
The center manifold theorem \cite{james2} which has been applied in \cite{JaNo,jamesnoble} states that, for $\mu ,\eta\approx 0$, small amplitude solutions of \eref{equv} (with $\left(u_{n},v_{n}\right)\in\mathbbm{D}$ for all $n\in \mathbb{Z}$) have the form
\begin{equation}
\label{solu}
\eqalign{
u_n = a_n \cos\left(t\right) + c_n \cos\left(3t\right) + \left[\varphi \left(a_n, d_n ,c_n ,b_n, \mu ,\eta \right)\right] \left(t\right),\\
v_n = d_n \cos\left(t\right) + b_n \cos\left(3t\right) + \left[\psi \left(a_n, d_n ,c_n ,b_n, \mu ,\eta \right)\right] \left(t\right).
}
\end{equation}
The functions $\left(\varphi ,\psi\right): \R^6 \rightarrow \mathbbm{D}$ are combinations of higher order Fourier components $\cos\left(\left(2k+1\right)t\right)$ with $k\geqslant 2$, they depend smoothly on $Y^c_n =\left(a_n, d_n ,c_n ,b_n\right)$, $\mu ,\eta$, and contain only higher order terms in $Y^c_n ,\mu ,\eta$, i.\,e.,
\begin{equation}
\label{deriv}
\varphi \left(0, 0 ,0 ,0, \mu ,\eta \right)=0,\qquad {\mathrm D}_{Y^c}\varphi \left(0, 0 ,0 ,0, \mu ,\eta \right)=0
\end{equation}
(the same holds for $\psi$). A four-dimensional center manifold, locally invariant under $ F_{m,\omega}$ and containing all small amplitude solutions, is locally constructed as the graph of the function $\left(\varphi ,\psi \right)\left[\cdot, \mu ,\eta \right]$ defined on the central subspace. As indicated by \eref{deriv}, the center manifold is tangent to the central subspace at $Y_n =0$ (even for $( \mu ,\eta )\neq 0$, since the harmonics $\cos\left(t\right)$ and $\cos\left(3t\right)$ span an invariant subspace of the linearized dynamics for all parameter values). 
 
The center manifold reduction theorem can be intuitively understood by considering the linearized system at $Y_n=0$, given by $Y_{n+1}={\mathrm D}F_{m,\omega}\left(0\right)\, Y_n$ for $\left(\omega,m\right)=\left(\frac{1}{2},\frac{1}{8}\right)$. In the linear case, all bounded solutions lie on the $4$-dimensional invariant central subspace (all the other solutions diverge exponentially as $n\rightarrow +\infty$ or $-\infty$ since they possess a component along an hyperbolic mode). A similar result holds true locally in the nonlinear case, where the central subspace is replaced by a $4$-dimensional invariant local center manifold containing all small amplitude solutions. 

We note that $\varphi \left(\cdot,\mu ,\eta \right)$ and $\psi \left(\cdot,\mu ,\eta \right)$ commute with $-I$ (the symbol $I$ denotes the identity operator) since the center manifold is invariant under the symmetry $-I$ of \eref{recrel}, which originates from the evenness of $V$. Consequently, $\varphi$ and $\psi$ are of order $\| Y^c \|^3$ as $Y^c \rightarrow 0$ in $\R^4$, where $\|\cdot\|$ denotes the Euclidean norm. In the following, we denote by $\phi =\left(\varphi ,\psi\right)$ the reduction function having the center manifold as its graph. Furthermore, we denote by $P_c$ the spectral projection on the central subspace,
\begin{equation}
P_c \Big(\sum_{k=0}^{+\infty} A_k\cos\left(\left(2k+1\right)t\right)\Big) = A_0\cos\left(t\right)+A_1 \cos\left(3t\right),
\end{equation}
and denote by $\mathcal{I}$ the isomorphism
\begin{equation}
\mathcal{I}\left(a,d,c,b\right)^{\mathsf T}=
(a \cos\left(t\right) + c \cos\left(3t\right),\,
d \cos\left(t\right) + b \cos\left(3t\right))^{\mathsf T}
\end{equation}
between $\mathbb{R}^4$ and the central subspace. For $\mu ,\eta\approx 0$, the sequences of coordinates $Y^c_n \in \mathbb{R}^4$ of small amplitude solutions on the center manifold are given by the four-dimensional recurrence relation 
\begin{equation}
\label{recred}
Y^c_{n+1}=f_{\mu ,\eta}\left(Y^c_n\right),
\end{equation}
where 
\begin{equation}
f_{\mu ,\eta}\left(Y^c\right)
=\mathcal{I}^{-1}P_c\, F_{m,\omega}\!\left(\mathcal{I}Y^c+\phi \left(Y^c ,\mu ,\eta\right)\right)
\end{equation}
for all $Y^c \in \mathbb{R}^4$ in a neighborhood of $0$, and where the parameters $m$, $\omega$ are defined by \eref{defparam}. Equation \eref{recred} is obtained by substituting Ansatz \eref{solu} into \eref{recrel} and projecting the resulting equation onto the central subspace. Note that \eref{recred} inherits the invariance of \eref{recrel} under $-I$. Moreover, the invariance of \eref{equv} under the symmetry $\left(u_n,v_{n}\right)\to\left(-u_{-n},-v_{-n+1}\right)$ induces the invariance $\left(a_n, d_n ,c_n ,b_n\right)\to\left(-a_{-n}, -d_{-n+1} ,-c_{-n} ,-b_{-n+1}\right)$ in the recurrence relation \eref{recred}. Since $\phi = \Or \left(\| Y^c \|^3 \right)$ as $Y^c \rightarrow 0$, we have in addition
\begin{equation}
\label{id2}
f_{\mu ,\eta}\left(Y^c\right)
=\mathcal{I}^{-1}P_c\, F_{m,\omega}\left(\mathcal{I}Y^c\right)
+\Or \left(\| Y^c \|^5 \right) .
\end{equation}
In what follows we compute the principal part of the reduced mapping \eref{recred}. Due to the complicated structure of $ F_{m,\omega}$, we shall not directly compute the Taylor expansion of \eref{id2} and proceed instead in the following way. Substituting Ansatz \eref{solu} into \eref{equv}, projecting onto the central subspace (i.\,e., applying $P_c$) and identifying the components along the harmonics $\cos\left(t\right)$ and $\cos\left(3t\right)$ results in the following recurrence relations (computations are lengthy but straightforward)
\numparts
\begin{eqnarray}
\fl d_{n+1}+d_n+\mu a_n=
-\tfrac{3}{4}\beta \left(d_{n}-a_n\right)\left[\left(d_{n}-a_n\right)^2+\left(d_{n}-a_n\right)\left(b_{n}-c_n\right)+2\left(b_{n}-c_n\right)^2\right]\nonumber\\
-\tfrac{3}{4}\beta \left(d_{n+1}-a_n\right)\left[\left(d_{n+1}-a_n\right)^2+\left(d_{n+1}-a_n\right)\left(b_{n+1}-c_n\right)+2\left(b_{n+1}-c_n\right)^2\right]\nonumber\\
+\Or \left(\left(\| Y^c_n \|+\| Y^c_{n+1} \|\right)^{5}\right),\label{cost1}\\
\fl b_{n+1}+b_n+\left(9 \mu + 16\right) c_n=
-\beta \left[\tfrac{1}{4}\left(d_{n}-a_n\right)^3+\tfrac{3}{2}\left(d_{n}-a_n\right)^2\left(b_{n}-c_n\right)+\tfrac{3}{4}\left(b_{n}-c_n\right)^3\right]\nonumber\\
-\beta \left[\tfrac{1}{4}\left(d_{n+1}-a_n\right)^3+\tfrac{3}{2}\left(d_{n+1}-a_n\right)^2\left(b_{n+1}-c_n\right)+\tfrac{3}{4}\left(b_{n+1}-c_n\right)^3\right] \nonumber\\
+\Or \left(\left(\| Y^c_n \|+\| Y^c_{n+1} \|\right)^{5}\right),\label{cos3t1}\\
\fl a_{n+1}+a_n+\left(\eta -\tfrac{7}{4}\right)d_{n+1}=\tfrac{3}{4}\beta \left(d_{n+1}-a_{n+1}\right)\nonumber\\
\times\left[\left(d_{n+1}-a_{n+1}\right)^2+\left(d_{n+1}-a_{n+1}\right)\left(b_{n+1}-c_{n+1}\right) +2\left(b_{n+1}-c_{n+1}\right)^2\right]\nonumber\\
+\tfrac{3}{4}\beta \left(d_{n+1}-a_n\right)\left[\left(d_{n+1}-a_n\right)^2+\left(d_{n+1}-a_n\right)\left(b_{n+1}-c_n\right)+2\left(b_{n+1}-c_n\right)^2\right]\nonumber\\
+\Or \left(\left(\| Y^c_n \|+\| Y^c_{n+1} \|\right)^{5}\right),\label{cost2}\\
\fl c_{n+1}+c_n+\left(\tfrac{1}{4}+9\eta \right)b_{n+1}=\nonumber\\
\beta \left[\tfrac{1}{4}\left(d_{n+1}-a_n\right)^3+\tfrac{3}{2}\left(d_{n+1}-a_n\right)^2\left(b_{n+1}-c_n\right)+\tfrac{3}{4}\left(b_{n+1}-c_n\right)^3\right]\nonumber\\
+\beta \left[\tfrac{1}{4}\left(d_{n+1}-a_{n+1}\right)^3+\tfrac{3}{2}\left(d_{n+1}-a_{n+1}\right)^2\left(b_{n+1}-c_{n+1}\right)+\tfrac{3}{4}\left(b_{n+1}-c_{n+1}\right)^3\right]\nonumber\\
+\Or \left(\left(\| Y^c_n \|+\| Y^c_{n+1} \|\right)^{5}\right).\label{cos3t2}
\end{eqnarray}
\endnumparts
Equations \eref{cost1}--\eref{cos3t2} represent an implicit form of the recurrence relation \eref{recred}. Using the implicit function theorem, one can reformulate \eref{cost1}--\eref{cos3t2} locally as a map $f_{\mu ,\eta}:\, \left(a_n , d_n, c_n, b_n\right)\mapsto \left(a_{n+1} , d_{n+1}, c_{n+1}, b_{n+1}\right)$.

In conclusion, in the parameter regime $\mu ,\eta\approx 0$, the center manifold theorem from \cite{james2} guarantees that small amplitude solutions of \eref{equv} are determined by a four-dimensional mapping whose principal part is given by \eref{cost1}--\eref{cos3t2} (in addition the reduced mapping preserves the symmetries of the infinite-dimensional system). Consequently one can state the following theorem.
\begin{theorem}
\label{reduction}
There exist neighbourhoods $\Omega$, ${\cal U}$ and $\Lambda$ of $\left(u,v\right)=0$ in $\mathbb{D}$, $Y^c =0$ in $\R^4$ and $\left(\mu ,\eta \right)=0$ in $\R^2$, respectively, and a $C^k$-map $\phi : \R^6 \rightarrow \mathbbm{D}$ with $\phi \left(0, 0 ,0 ,0, \mu ,\eta \right)=0$, ${\mathrm D}_{Y^c}\phi \left(0, 0 ,0 ,0, \mu ,\eta \right)=0$ such that the following holds for all $\left(\mu ,\eta \right)\in \Lambda$:
\begin{enumerate}
\item All solutions of \eref{equv} such that $\left(u_n,v_{n}\right)\in \Omega$ for all $n\in \Z$ have the form \eref{solu}, where $Y^c_n =\left(a_n, d_n ,c_n ,b_n\right)$ satisfies a $4$-dimensional recurrence relation with principal part given by \eref{cost1}--\eref{cos3t2}.
\item If $Y^c_n =\left(a_n, d_n ,c_n ,b_n\right)$ is a solution of \eref{cost1}--\eref{cos3t2} with $Y^c_n \in {\cal U}$ for all $n\in \Z$, then \eref{solu} defines a solution of \eref{equv}.
\item The invariance of \eref{equv} under the symmetries $\left(u_n,v_{n}\right)\to\left(-u_n,-v_{n}\right)$ and $\left(u_n,v_{n}\right)\to\left(-u_{-n},-v_{-n+1}\right)$ induces the invariance under $Y^c\to-Y^c$ and $\left(a_n, d_n ,c_n ,b_n\right)\to\left(-a_{-n}, -d_{-n+1} ,-c_{-n} ,-b_{-n+1}\right)$ for the reduced system \eref{cost1}--\eref{cos3t2}.
\end{enumerate}
\end{theorem}

Small amplitude orbits of the reduced map homoclinic to $0$ correspond via theorem \ref{reduction} to exact discrete breather solutions of the FPU system. In what follows we shall not prove the existence of such exact homoclinic orbits. Instead we shall derive in the next section a continuum limit differential system which provides approximate homoclinic orbits of the map. These analytical results will be used in section \ref{sec:comparison} to explain the numerically observed breather bifurcations.

In addition, the leading order continuum limit of the reduced map obtained in the next section might be used, together with reversibility [theorem \ref{reduction} (iii)], as a tool for proving the existence of exact small amplitude homoclinics close to the continuum limit.

\vspace{2ex}

\section{Continuum limit and homoclinic bifurcations for the reduced map}
\label{sec:asymptotics}

\subsection{Continuum limit of the reduced map}

We assume, as before, that $\mu ,\eta \approx 0$ with $\mu >0$, and introduce a small parameter $\epsilon$, setting $\mu = \epsilon^2$ and $\eta = \kappa \epsilon^2$. For $\epsilon =0$, the recurrence relation \eref{cost1}--\eref{cos3t2} linearized at the fixed point $\left(a,d,c,b\right)=0$ reads
\numparts
\begin{eqnarray}
0=d_{n+1}+d_n,\\
0=b_{n+1}+b_n+16 c_n,\\
0=a_{n+1}+a_n-\tfrac{7}{4}d_{n+1},\\
0=c_{n+1}+c_n+\tfrac{1}{4}b_{n+1},
\end{eqnarray}
\endnumparts
with bounded solutions given by
\begin{equation}
\label{line}
\left(a_n , d_n, c_n, b_n\right)^{\mathsf T}=a \left(\left(-1\right)^n,0,0,0\right)^{\mathsf T} + b\, \left(0,0,-\tfrac{1}{8},1\right)^{\mathsf T},
\end{equation}
$a,b\in \mathbb{R}$ being arbitrary. Indeed, for $\epsilon =0$ the spectrum of ${\mathrm D}F_{m,\omega}\left(0\right)$ on the unit circle consists in a double non-semi-simple eigenvalue $-1$ [with the corresponding invariant subspace spanned by $\cos\left(t\right)$] and a double non-semi-simple eigenvalue $+1$ [invariant subspace spanned by $\cos\left(3t\right)$]. 

We search for small amplitude approximate solutions of \eref{cost1}--\eref{cos3t2} in the form of a slow modulation of \eref{line} supplemented by higher harmonics, i.\,e.,
\begin{equation}
\label{expansion}
\fl
\left(\!
\begin{array}{c}
a_n \\
d_{n+1} \\
c_n \\
b_{n+1}
\end{array}
\!\right)
=
\epsilon A_0\left(\xi \right)
\left(\!
\begin{array}{c}
\left(-1\right)^n \\
0 \\
0 \\
0
\end{array}
\!\right)
+ 
\epsilon B_0\left(\xi \right)
\left(\!
\begin{array}{c}
0 \\
0 \\
-\tfrac{1}{8} \\
1
\end{array}
\!\right)
+
\sum_{k\geqslant 2}{\epsilon^k \left(\left(-1\right)^n V_k\left(\xi \right)+ \widetilde{V}_k\left(\xi \right)\right)}
\end{equation}
where $\xi =\epsilon n$. This approach has some analogy with the multiple scales expansion technique (see \cite{gian3} and its references), with the fundamental difference that the FPU system has been locally reduced for time-periodic solutions (by theorem \ref{reduction}) to a finite-dimensional problem. In what follows we compute the principal terms in the expansion \eref{expansion}. Inserting \eref{expansion} into the recurrence relations \eref{cost1}--\eref{cos3t2} and identifying the different harmonics and powers of $\epsilon$, we obtain
\numparts
\begin{eqnarray}
\label{an1l}
a_n & = & \left(-1\right)^n \epsilon A_0\left(\epsilon n\right)+\Or(\epsilon^2),\\
b_{n} & = & \epsilon B_0\left(\epsilon n\right)+\Or(\epsilon^2),\\
c_n & = & -\tfrac{1}{8}\epsilon B_0\left(\epsilon n\right)+\Or(\epsilon^2),\\
\label{an4l}
d_{n} & = & \tfrac{4}{7}\left(-1\right)^{n} \epsilon^2 \frac{\rmd A_0}{\rmd\xi}\left(\epsilon n\right) +\Or(\epsilon^3),
\end{eqnarray}
\endnumparts
where $A_0,B_0$ are solutions of system
\numparts
\begin{eqnarray}\label{eqal}
0=\frac{\rmd^2A_0}{\rmd\xi^2}
-\tfrac{7}{4}A_0
+\tfrac{21}{8}\beta \left(
A^3_0
+\tfrac{81}{32}A_0B^2_0
\right),\\
\label{eqbl}
0=\frac{\rmd^2B_0}{\rmd\xi^2}
-9\left(\tfrac{1}{4}+16\kappa \right)B_0
+\tfrac{243}{8}\beta \left(
2A^2_0 B_0
+\tfrac{81}{64}B^3_0
\right).
\end{eqnarray}
\endnumparts
Details of this calculation are reported in the appendix. In this way we have obtained a formal continuum limit \eref{eqal}--\eref{eqbl}, valid for $\mu,\eta\approx0$, of the four-dimensional recurrence relation \eref{cost1}--\eref{cos3t2}. The corresponding approximate solutions \eref{an1l}--\eref{an4l} of the reduced recurrence relation in turn correspond to approximate solutions of the FPU system \eref{equv} having the form
\begin{equation}\label{solul}
\eqalign{
u_n = \left(-1\right)^n \epsilon A_0\left(\epsilon n\right)\cos\left(t\right)-\tfrac{1}{8}\epsilon B_0\left(\epsilon n\right)\cos\left(3t\right)+\Or(\epsilon^2),\\
v_n = \epsilon B_0\left(\epsilon n\right)\cos\left(3t\right)
+\Or(\epsilon^2) .
}
\end{equation}

Normalizing the solutions of the system \eref{eqal}--\eref{eqbl} by setting $B_0\left(\xi\right) =\frac{8}{\sqrt{7}}b\left(\xi\right)$ and $A_0\left(\xi\right)=a\left(\xi\right)$, the equations read
\numparts
\begin{eqnarray}
\label{eqar}
\frac{\rmd^2a}{\rmd\xi^2}-\tfrac{7}{4}a+\tfrac{7}{27}h\, a^3+6h\, ab^2=0,\\
\label{eqbr}
\frac{\rmd^2b}{\rmd\xi^2}-9\left(\tfrac{1}{4}+16\kappa\right)b+6h\,a^2b+\tfrac{243}{7}h\,b^3=0,
\end{eqnarray}
\endnumparts
with
\begin{equation}
h=\tfrac{81}{8}\beta.
\end{equation}

In this notation, approximate solutions, valid for $\mu,\eta\approx0$, of the original equations of motion \eref{fdintro} of the diatomic FPU chain are given by 
\begin{equation}
\label{solpair}
\eqalign{
X_{2n}\left(t\right) = \left(-1\right)^n \epsilon a\left(\epsilon n\right)\cos\left(\omega t\right)-\tfrac{1}{\sqrt{7}}\epsilon b\left(\epsilon n\right) \cos\left(3\omega t\right) +\Or(\epsilon^2),\\
X_{2n-1}\left(t\right) = 
\epsilon \tfrac{8}{\sqrt{7}}b\left(\epsilon n\right)\cos\left(3\omega t\right)
+\Or(\epsilon^2).
}
\end{equation}
The small parameter $\epsilon$ (which measures the amplitude of the solutions), the frequency $\omega$, the mass ratio $m$ and the parameter $\kappa$ in \eref{eqbr} are linked by the relations
\begin{equation}
\label{params}
\frac{\omega^2}{m} = 2+\epsilon^2,\qquad
\omega^2 =\frac{1}{4}+ \kappa \epsilon^2.
\end{equation}
The asymptotic analysis leading to this result is {\em a priori}\/ valid for $\epsilon \approx 0$, $\kappa$ being fixed. This corresponds to the regime $\omega\approx \frac{1}{2}$, $m\approx \frac{1}{8}$, with $m/\omega^2 < \frac{1}{2}$.

\subsection{Homoclinic bifurcations in the continuum limit}

In this section we analyze some local homoclinic bifurcations for system \eref{eqar}--\eref{eqbr}. These bifurcations show up for hard interaction potentials, i.\,e., we shall assume $\beta >0$ in the potential $V$ in \eref{eq:V2}. 

Homoclinic bifurcations in systems of the type \eref{eqar}--\eref{eqbr} have been analyzed by several authors in different contexts (see e.\,g.\ \cite{yang,boyd,champneys}). For example, such a system can be obtained when considering two coupled nonlinear Schr\"odinger equations (CNLS system) and one is looking for time-periodic solution components. The CNLS system has been introduced to describe coupled mode dynamics in different physical contexts, such as fluid dynamics or nonlinear optics (see e.g. references in \cite{boyd}). In the context of nonlinear lattices, the CNLS system has been derived in reference \cite{Konotop} for the Kac-Baker model, in order to describe the nonlinear modulation of two superposed linear modes with equal group velocities (this analysis is based on a formal multiscale expansion). 

Here we consider pitchfork bifurcations from a single-component homoclinic orbit of \eref{eqar}--\eref{eqbr} ($\left(a,0\right)$ or $\left(0,b\right)$) to two-components homoclinic orbits $\left(a,b\right)$. These bifurcations have been analyzed by Yang \cite{yang} using a classical perturbative analysis for pitchfork bifurcations (see e.\,g.\ \cite{ioossjos}), and the computations in reference \cite{yang} can be readily employed here. As will be shown in the next section, these bifurcations correspond (qualitatively and quantitatively) to the numerically found breather bifurcations in section \ref{sec:cornerbifu}.

\subsubsection{Homoclinic orbits bifurcating from $\left(0,b\right)$:}

We first normalize the coefficients of \eref{eqar}--\eref{eqbr}. We assume $\kappa > -\frac{1}{64}$ and set
\begin{equation}
b\left(\xi \right)=a_0\,  u\left(\delta \xi\right),\qquad
a\left(\xi \right)=a_0\,  v\left(\delta \xi\right),
\end{equation}
where $a_0 = \left[ \, \frac{7}{27h}\left(\frac{1}{4}+16\kappa \right)\, \right]^{1/2}$ and $\delta = 3\, \left(\frac{1}{4}+16\kappa \right)^{1/2}$. This yields
\begin{equation}
\label{equr}
\eqalign{
\frac{\rmd^2 u}{\rmd x^2}-u+u^3+{\beta_0} u v^2=0,\\
\frac{\rmd^2 v}{\rmd x^2}-\Omega^2 v+{\beta_0} u^2 v+\gamma v^3=0,
}
\end{equation}
where $x=\delta \xi$, ${\beta_0}=\frac{14}{81}$, $\gamma = \left(\frac{7}{81}\right)^2$ and $\Omega =\frac{\sqrt{7}}{6} \, \left(\frac{1}{4}+16\kappa \right)^{-1/2}$. System \eref{equr} has the homoclinic solution $\left(u,v\right)=\left(u_h,0\right)$ with $u_h\left(x\right)=\sqrt{2}\, \mbox{sech}\left(x\right)$. We shall look for critical values of $\Omega$ at which solutions homoclinic to $0$ bifurcate from $\left(u,v\right)=\left(u_h,0\right)$.

This problem has been previously analyzed by Yang \cite{yang} for a slightly more symmetric system (with $\gamma =1$). This analysis readily applies here (with only a few straightforward modifications of the coefficients) and we refer the reader to \cite{yang} for details (see also \cite{champneys} for more details on the linearized problem at $\left(u,v\right)=\left(u_h,0\right)$). There exists a finite number of critical values 
\begin{equation}
\Omega_n = s-n,
\end{equation}
where $s=\frac{1}{2}\left(\sqrt{1+8\beta_0 }-1\right)$ and $n\geqslant 0$ is an integer. A pitchfork bifurcation occurs from $\left(u,v\right)=\left(u_h,0\right)$ at any of these critical values, due to the invariance $v\rightarrow -v$ of \eref{eqar}--\eref{eqbr}. If $n$ is even, bifurcating solutions are even in $x$. If $n$ is odd, $u$ is even and $v$ is odd for bifurcating solutions. The number of critical values depends on the size of $\beta_0$, and is equal to one in our case. We have $\Omega_0 \approx 0.272$ which corresponds to $\kappa = \kappa_0 \approx 0.149$.

The branch of homoclinics bifurcating from $\left(u,v\right)=\left(u_h,0\right)$ as $\Omega \approx \Omega_0$ can be locally parameterized by
\begin{eqnarray}
\label{bifpt1}
u&=&\sqrt{2}\, \mbox{sech}\left(x\right) + \Or\left(\lambda^2 \right),\\
\label{bifpt2}
v&=&\lambda \, \mbox{sech}^s\left(x\right)+ \Or\left(|\lambda |^3 \right),\\
\label{bifpt3}
\Omega^2& =& \Omega^2_0 + \theta\, \lambda^2 + \Or\left(\lambda^4 \right),
\end{eqnarray}
where $\lambda$ is a small parameter (see \cite{yang}, paragraph 3.1). The sign of
\begin{equation}
\theta = \left(\gamma - s^3\right) \frac{ \int_{\mathbb{R}}{\mbox{sech}^{4s}\left(x\right)\, \rmd x} }{ \int_{\mathbb{R}}{\mbox{sech}^{2s}\left(x\right)\, \rmd x}  }
\end{equation}
determines whether the pitchfork bifurcation is sub- or supercritical. Here $\gamma -s^3\approx -0.01$ and consequently one has $\Omega < \Omega_0$ and $\kappa > \kappa_0$ for bifurcating homoclinic solutions. Returning to the original parameters $\left(\omega^2 ,m\right)$ given by \eref{params} and fixing $\omega^2 > 1/4$, bifurcating homoclinics exist for 
\begin{equation}
\label{critpt}
m>m_{\mathrm{PT}}\left(\omega \right)=\tfrac{1}{8}+\alpha_{\mathrm{PT}}\left(\omega^2 -\tfrac{1}{4}\right)+\Or\left[\left(\omega^2 -\tfrac{1}{4}\right)^2 \right], 
\end{equation}
where $\alpha_{\mathrm{PT}} = \frac{1}{2}-\frac{1}{16 \kappa_0}\approx 0.08$, and the index PT is for ``period tripling''.

\subsubsection{Homoclinic orbits bifurcating from $\left(a,0\right)$:}

We again normalize the coefficients of \eref{eqar}--\eref{eqbr}. We assume $\kappa > -\frac{1}{64}$ and set
\begin{equation}
a\left(\xi \right)=a_1 \, u\left(\tfrac{\sqrt{7}}{2} \xi\right),\qquad
b\left(\xi \right)=a_1 \, v\left(\tfrac{\sqrt{7}}{2} \xi\right),
\end{equation}
where $a_1 = \left( \, \frac{27}{4h}\, \right)^{1/2}$. This yields
\begin{equation}
\eqalign{
\label{equrt}
\frac{\rmd^2 u}{\rmd x^2}-u+u^3+{\widetilde\beta_0}uv^2=0,\\
\frac{\rmd^2 v}{\rmd x^2}-\widetilde\Omega^2 v+{\widetilde\beta_0}u^2v+\widetilde\gamma v^3=0,
}
\end{equation}
where $x=\frac{\sqrt{7}}{2} \xi$, ${\widetilde\beta_0}=\frac{162}{7}$, $\widetilde\gamma = \left(\frac{81}{7}\right)^2$ and $\widetilde\Omega =\frac{6}{\sqrt{7}} \left(\frac{1}{4}+16\kappa \right)^{1/2}$. System \eref{equrt} is of the same type as \eref{equr}, hence the analysis performed in the previous section still applies. We look for critical values of $\widetilde\Omega$ at which solutions homoclinic to $0$ bifurcate from $\left(u,v\right)=\left(u_h,0\right)$. These critical values are given by 
\begin{equation}
\widetilde\Omega_n = \widetilde{s}-n,
\end{equation}
where $\widetilde{s}=\frac{1}{2}\left(\sqrt{1+8\widetilde\beta_0 }-1\right)\approx 6.322$ and $n$ is an integer with $0\leqslant n\leqslant 6$.

In what follows, we concentrate on the bifurcation near $\widetilde\Omega_0 = \widetilde{s}$, which corresponds to $\kappa = \widetilde\kappa_0 \approx 0.47$. The branch of homoclinics bifurcating from $\left(u,v\right)=\left(u_h,0\right)$ as $\widetilde\Omega \approx \widetilde\Omega_0$ can be locally parameterized by
\begin{eqnarray}
\label{bifac1}
u&=&\sqrt{2}\, \mbox{sech}\left(x\right) + \Or\left(\lambda^2 \right),\\
\label{bifac2}
v&=&\lambda \, \mbox{sech}^{\widetilde{s}}\left(x\right)+ \Or\left(|\lambda |^3 \right),\\
\label{bifac3}
\widetilde\Omega^2& =& \widetilde\Omega^2_0 + \widetilde\theta\, \lambda^2 + \Or\left(\lambda^4 \right),
\end{eqnarray}
where $\lambda$ is a small parameter. Since $\widetilde\theta$ has the same sign as $\widetilde\gamma - \widetilde{s}^{\,3}$, one has $\widetilde\theta <0$ and $\widetilde\Omega < \widetilde\Omega_0$ for bifurcating homoclinic solutions. Therefore $\kappa < \widetilde\kappa_0$ on the bifurcating solution branch. Returning to the original parameters $\left(\omega^2 ,m\right)$ given by \eref{params} and fixing $\omega^2 > 1/4$, bifurcating homoclinics exist for 
\begin{equation}
\label{critptbis}
m<m_{\mathrm{HM}}\left(\omega \right)=\tfrac{1}{8}+\alpha_{\mathrm{HM}}\left(\omega^2 -\tfrac{1}{4}\right)+\Or\left[\left(\omega^2 -\tfrac{1}{4}\right)^2 \right], 
\end{equation}
where $\alpha_{\mathrm{HM}} = \frac{1}{2}-\frac{1}{16 \widetilde\kappa_0}\approx 0.367$, and the index HM is for ``heavy mass oscillations''.

We end with a remark concerning the above pitchfork bifurcations in \eref{eqar}--\eref{eqbr}, related with the invariances $\left(a,b\right)\rightarrow \left(-a,b\right)$ and $\left(a,b\right)\rightarrow \left(a,-b\right)$. This system formally corresponds to \eref{cost1}--\eref{cos3t2} in a continuum limit, but the corresponding symmetries $\left(a_n,d_n,c_n,b_n\right)\rightarrow \left(-a_n,-d_n,c_n,b_n\right)$ and $\left(a_n,d_n,c_n,b_n\right)\rightarrow \left(a_n,d_n,-c_n,-b_n\right)$ are lost for the exact system \eref{cost1}--\eref{cos3t2}. Consequently, we only expect imperfect pitchfork bifurcations (i.\,e., with nonsymmetric bifurcating solution branches) when dealing with the exact reduced map.

\section{Approximate breather solutions and comparison with numerical results}
\label{sec:comparison}

In this section, the discrete breathers observed numerically in section \ref{sec:numresults} are shown to correspond to the different branches of solutions of the system of equations \eref{eqar}--\eref{eqbr} discussed in the previous section.

Assuming the case of a {hard potential} where $\beta >0$ in the potential \eref{eq:W}, system \eref{eqar}--\eref{eqbr} admits homoclinic solutions to $0$ with $b=0$, and possesses homoclinic solutions to $0$ with $a=0$ if $\kappa >-\frac{1}{64}$.

Homoclinic solutions of \eref{eqar}--\eref{eqbr} with $a=0$ and $b\left(\xi \right)=\sqrt{2}\, a_0\, \mbox{sech}\left(\delta \xi \right)$ correspond to breather solutions with frequencies slightly above the optic band. The principal part of these solutions as the mass ratio $m$ is close to $\frac{1}{8}$ is provided by \eref{solpair}. Heavy and light masses oscillate at the same frequency $3\omega$, and light mass displacements are larger than heavy mass displacements (see figure \ref{fig:RWAshapeM<=>m}, bottom row). The existence of these solutions has been proved in \cite{jamesnoble} for an arbitrary mass ratio $m$ in the small amplitude limit (for frequencies slightly above the optic band). Among the solution families obtained in \cite{jamesnoble}, one family has the symmetry $X_{-n}\left(t\right)=X_n \left(t\right)$ for even potentials, whereas the principal part of \eref{solpair} only provides $X_{-n}\left(t\right)-X_n \left(t\right)=\Or \left(\epsilon^2\right)$. To better describe this symmetry, it is more convenient to rewrite \eref{solpair} in the form
\begin{equation}
\label{opticpair}
\eqalign{
X_{2n}\left(t\right) =
-\sqrt{\tfrac{2}{7}}\, a_0 \epsilon \,\textnormal{sech}\left(\delta \epsilon n\right) \cos\left(3\omega t\right)+\Or(\epsilon^2),\\
X_{2n-1}\left(t\right) =
\tfrac{8}{\sqrt{7}}\, a_0 \epsilon \, \textnormal{sech}\left[\delta \epsilon \left(n-\tfrac{1}{2}\right)\right] \cos\left(3\omega t\right)+\Or(\epsilon^2),
}
\end{equation}
valid for fixed $\kappa > -\frac{1}{64}$ and $\epsilon \approx 0$ (see figure \ref{fig:RWAshapeM<=>m}, bottom, for a comparison of this asymptotic expression with a numerically computed discrete breather solution). Fixing $\kappa$ and varying $\epsilon$ is equivalent to moving parameters on the curve $\Gamma_\kappa$ defined by
\begin{equation}
\frac{1}{\kappa}\left(\omega^2-\frac{1}{4}\right)=\frac{\omega^2}{m}-2.
\end{equation}
All these curves pass through $\left(\omega^2,m\right)=\left(\frac{1}{4},\frac{1}{8}\right)$. The first limiting curve $\Gamma_{-1/64}$ is tangent at this point to the line $9\omega^2=2\left(1+m\right)$ (where $3\omega$ reaches the top of the optic band), and the second limiting curve $\Gamma_{\infty}$ corresponds to the line $ \omega^2=2m$ (top of the acoustic band).

Homoclinic solutions of \eref{eqar}--\eref{eqbr} with $b=0$ and $a\left(\xi \right)=\sqrt{2}\, a_1\, \mbox{sech}\left(\frac{\sqrt{7}}{2} \xi\right)$ correspond to gap breather solutions with frequencies slightly above the acoustic band. Heavy and light masses oscillate with the same frequency $\omega$ and heavy mass displacements are larger than light mass displacements (see figure \ref{fig:RWAshapeM<=>m}, top). The existence of such solutions has been proved in \cite{jamesnoble} in the small amplitude limit (for frequencies slightly above the acoustic band), and for mass ratio $m$ outside intervals of resonance in which the acoustic band and an image of the optic band intersect (see shaded areas in figure \ref{bands}). As observed in section \ref{sec:loopspace}, for an even interaction potential only the odd multiples of the optic band constitute forbidden zones. The mass ratio $m=\frac{1}{8}$ precisely belongs to the boundary of one of these zones. System \eref{solpair} formally extends the set of solutions obtained in \cite{jamesnoble}, since it provides approximate acoustic breather solutions with parameters close to $\left(\omega^2,m\right)=\left(\frac{1}{4},\frac{1}{8}\right)$, $\kappa > -\frac{1}{64}$ being fixed and $\epsilon$ close to $0$,
\begin{equation}
\label{acousticpair}
\eqalign{
X_{2n}\left(t\right) =
\left(-1\right)^n \sqrt{2}\, a_1 \epsilon\, \textnormal{sech}\left(\tfrac{\sqrt{7}}{2} \epsilon n\right) \cos\left(\omega t\right)+\Or(\epsilon^2),\\
X_{2n-1}\left(t\right) =\Or(\epsilon^2)
}
\end{equation}
(see figure \ref{fig:RWAshapeM<=>m}, top, for a comparison of this asymptotic expression with a numerically computed discrete breather solution).

Homoclinic solutions with two nonvanishing components $a,b$ mix the two types of breather solutions (see figure \ref{fig:RWAshapeM<=>m}, middle). According to \eref{solpair}, a period tripling bifurcation occurs when a solution with a nonvanishing $a$-component emerges from a solution consisting of a $b$-component only (figure \ref{fig:RWAshapeM<=>m}, from bottom to top). The case when a $b$-component emerges from a solution consisting of an $a$-component only does not produce a frequency change but it modifies the spatial pattern of light mass oscillations and increases their amplitude. 

Let us now provide approximate expressions for the additional discrete breather solutions near the period tripling bifurcation.

The bifurcating homoclinics defined by \eref{bifpt1}--\eref{bifpt3} and the approximation \eref{solpair} provide approximate solutions 
\begin{equation}
\label{opticpairbif}
\eqalign{
\fl X_{2n}\left(t\right)=
-\sqrt{\tfrac{2}{7}} a_0 \epsilon\, \textnormal{sech}\left(\delta \epsilon n\right) \cos\left(3\omega t\right)\!+\!\left(-1\right)^n\! a_0 \epsilon \lambda\, \textnormal{sech}^s \left(\delta \epsilon n\right)\cos\left(\omega t\right)\!+\!\Or \!\left(\epsilon^2 \!+\!\epsilon \lambda^2 \right)\!, \\
\fl X_{2n-1}\left(t\right)=
\tfrac{8}{\sqrt{7}} a_0 \epsilon \,\textnormal{sech}\left[\delta \epsilon \left(n-\tfrac{1}{2}\right)\right] \cos\left(3\omega t\right)+\Or \left(\epsilon^2 +\epsilon \lambda^2 \right),
}
\end{equation}
parameterized by $\lambda \approx 0$ and $\epsilon \ll \lambda$, with
\begin{eqnarray}
\epsilon^2 = \frac{\omega^2}{m}-2,\\
\kappa - \kappa_0 = -\tfrac{9}{28}\, \theta\, \left(\tfrac{1}{4}+16\kappa_0 \right)^2 \lambda^2 + 
\Or(\lambda^4),\\
\omega^2 = \tfrac{1}{4}+\kappa \epsilon^2 .
\end{eqnarray}
In figure \ref{fig:RWAshapeM<=>m}, third row, we compare this asymptotic expression for $\epsilon >0$ and $\lambda >0$ (right column) with a numerically computed discrete breather solution (middle column). The numerically computed discrete breather in the left column would correspond to $\epsilon >0$ and $\lambda <0$. Note that the map $\left(\lambda^2 , \epsilon^2 \right)\mapsto \left(\omega^2 ,m\right)$ is locally invertible, and the value of $\left(\lambda , \epsilon \right)$ is determined by the choice of $\left(\omega^2 ,m\right)$ specified in the figure. Moreover, bifurcating homoclinics with $\lambda \neq 0$ locally exist for parameters above a curve $m=m_{\mathrm{PT}}\left(\omega \right)$ passing through $\left(\omega^2,m\right)=\left(\frac{1}{4},\frac{1}{8}\right)$, with the local approximation \eref{critpt}.

\begin{figure}
\center
\includegraphics[width=8cm,height=8cm,clip=false]{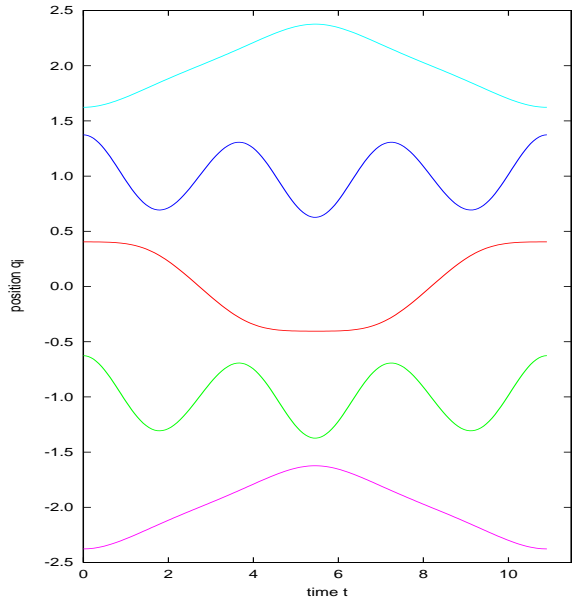}
\caption{\label{fig:osci} Mass displacements as a function of time, plotted over one oscillation period for the five central particles of a gap breather close to an order three frequency resonance between the optic phonon with wave number $k=0$ and the acoustic phonon with wave number $k=\pi$. The small masses (second and fourth curve from above) oscillate at leading order with three times the frequency of the heavy ones. Higher order corrections account for the slight modulation visible in the small masses' trajectories.}
\end{figure}

The local bifurcation around acoustic breathers can be described in a similar way. The bifurcating homoclinics defined by \eref{bifac1}--\eref{bifac3} and the approximation \eref{solpair} provide approximate solutions 
\begin{equation}
\label{acousticpairbif}
\eqalign{
\fl X_{2n}\left(t\right)=
\left(-1\right)^n  \sqrt{2}\, a_1 \epsilon\, \textnormal{sech}\left(\tfrac{\sqrt{7}}{2} \epsilon n\right) \cos\left(\omega t\right)-\tfrac{1}{\sqrt{7}}\, a_1 \epsilon \lambda\, \textnormal{sech}^{\widetilde{s}}\left(\tfrac{\sqrt{7}}{2} \epsilon n\right)
\cos\left(3\omega t\right)\\
+\Or \left(\epsilon^2 + \epsilon \lambda^2 \right),\\
\fl X_{2n-1}\left(t\right)=
\tfrac{8}{\sqrt{7}}\, a_1 \epsilon \lambda\, \textnormal{sech}^{\widetilde{s}}\left[\tfrac{\sqrt{7}}{2} \epsilon \left(n-\tfrac{1}{2}\right)\right]
\cos\left(3\omega t\right)+\Or \left(\epsilon^2 +\epsilon \lambda^3 \right),
}
\end{equation}
parameterized by $\lambda \approx 0$ and $\epsilon \ll \lambda$, with
\begin{eqnarray}
\epsilon^2 = \frac{\omega^2}{m}-2,\\
\kappa - \widetilde\kappa_0 =\tfrac{7}{576}\, \widetilde\theta \lambda^2 + \Or \left(\lambda^4 \right),\\
\omega^2 =\tfrac{1}{4}+\kappa \epsilon^2
\end{eqnarray}
(see figure \ref{fig:RWAshapeM<=>m}, second row, for a comparison of this asymptotic expression with a numerically computed discrete breather solution, in the case $\epsilon , \lambda >0$). Moreover, bifurcating homoclinics with $\lambda \neq 0$ locally exist for parameters below a curve $m=m_{\mathrm{HM}}\left(\omega \right)$ passing through $\left(\omega^2,m\right)=\left(\frac{1}{4},\frac{1}{8}\right)$, with the local approximation \eref{critptbis}. At leading order in $\epsilon$ and $\lambda$, the frequency of light masses is three times the frequency $\omega$ of heavy masses. This behavior corresponds to what is observed numerically for discrete breathers in this region of parameter space (see figure \ref{fig:osci} for an illustration). 

Comparing the numerically computed discrete breathers with the analytic ap\-prox\-i\-ma\-tions \eref{opticpair}, \eref{acousticpair}, \eref{opticpairbif}, and \eref{acousticpairbif} obtained by means of a center manifold reduction as plotted in figure \ref{fig:RWAshapeM<=>m}, we found very good {\em qualitative}\/ agreement of the results. Furthermore, excellent agreement is found when comparing the asymptotic behaviour of the bifurcation lines \eref{critpt} and \eref{critptbis} of the reduced system \eref{eqar}--\eref{eqbr} with the numerically computed bifurcation lines close to the ``corner'' formed by the phonon bands (see figure \ref{fig:bifudiagram}). The {\em quantitative}\/ agreement of the numerically computed discrete breathers with their analytic ap\-prox\-i\-ma\-tions depends, as expected, on how good the assumptions of the asymptotic analysis are met. For the weakly localized breathers in the upper row of figure \ref{fig:RWAshapeM<=>m}, the oscillation amplitudes from numerical and analytic computations agree very well, whereas for the stronger localized breathers plotted in the bottom row the amplitudes differ by a factor of approximately $\frac{2}{3}$. Considering parameter values $\left(\omega^2,m\right)$ closer to the phonon band edges, the localization strength of these breathers should decrease, leading to an improved agreement of numerical and analytic results. Unfortunately, we encountered problems with the convergence of the algorithm used for the numerical computation of the discrete breathers which prevented us from making such a comparison.

\section{Summary and conclusions}
\label{sec:summary}

Discrete breathers in a diatomic Fermi-Pasta-Ulam chain have been investigated by means of numerical as well as analytical methods. The technique of numerical continuation, based on Newton's method for computing zeros of a function, was employed for the numerical analysis. Bifurcations between different types of discrete breathers were found, with the mass ratio $m$ and the breather frequency $\omega$ as bifurcation parameters. A common feature of the bifurcation lines observed is that they emerge from corners or edges formed by phonon bands and their images in the parameter plane $\left(\omega^2,m\right)$.

Above the optic band, a bifurcation between symmetric and asymmetric discrete breathers was investigated. The bifurcation line, starting from the upper edge of the optic band at mass ratio $m=1$, was determined.

In the gap between the acoustic and the optic band, discrete breathers are believed to exist only outside certain ``images'' of the optic band, i.\,e., when their frequencies are nonresonant with the optic phonons. These images cross the upper band edge of the acoustic band, forming ``corners'' in parameter space. We chose a particular corner at $\left(\omega^2,m\right)=\left(\frac{1}{4},\frac{1}{8}\right)$ and investigated small amplitude discrete breathers in its vicinity. Two types of bifurcations, with bifurcation lines approaching the corner, were observed in this region. One is a period tripling bifurcation, the other is a local bifurcation around an acoustic breather increasing light masses oscillations with respect to heavy masses oscillations. Varying parameters across these two bifurcations, we have computed new types of breather solutions interpolating between the known optic and acoustic breathers.

Using the center manifold reduction theorem, we have proved locally [for $\left(\omega^2,m\right)\approx \left(\frac{1}{4},\frac{1}{8}\right)$] that all small amplitude (time-reversible) periodic oscillations are determined by a four-dimensional mapping. Homoclinic solutions of this reduced system have been studied formally by passing to a continuum limit, and a system of differential equations was derived, providing exact asymptotic expressions for discrete breathers close to $\left(\omega^2,m\right)=\left(\frac{1}{4},\frac{1}{8}\right)$. These expressions allow to identify the different types of breathers observed numerically with different branches of solutions of the set of equations. This analytic result confirms that, as suggested by the numerical data, the bifurcation lines emanate from $\left(\omega^2,m\right)=\left(\frac{1}{4},\frac{1}{8}\right)$, i.\,e., at the corner formed by the acoustic band and an image of the optic band. Moreover we have analytically computed the slope of the two bifurcation lines at this point.

We want to conclude with remarks concerning the above results. 

Although numerical as well as analytic results were derived in the vicinity of a particular crossing of upper edge of the acoustic band with that of the second image of the optic band at $\left(\omega^2,m\right)=\left(\frac{1}{4},\frac{1}{8}\right)$, similar bifurcations should take place close to all the similar ``corners'' in parameter space (with the harmonic $\cos{\left(3t\right)}$ in \eref{solul} replaced by an harmonic $\cos{\left(\left(k+1\right)t\right)}$ near the $k$-th image of the optic band). Moreover one should be able to detect $5$ additional bifurcation curves near $\left(\omega^2,m\right)=\left(\frac{1}{4},\frac{1}{8}\right)$, corresponding to the additional critical values $\widetilde\Omega_1 ,\ldots ,\widetilde\Omega_5$ leading to bifurcations around acoustic breathers (we have not explored this problem numerically).

Lastly, an interesting result would be to prove for the exact map \eref{cost1}--\eref{cos3t2} the existence of homoclinic orbits to $0$ close to the approximate homoclinics of section \ref{sec:asymptotics}. This would automatically imply by theorem \ref{reduction} the existence of corresponding exact breather solutions for the FPU system.

\ack
We are grateful to Roberto Livi for suggesting collaboration and initiating research reported in this article. M.~K.\ would like to thank Francesco Piazza for providing ``seeds'' for the numerical computation from rotating wave approximation. This work was partially supported by EU contract HPRN-CT-1999-00163 (LOCNET network).

\appendix
\section*{Appendix}
\setcounter{section}{1}

In this appendix we report the details of the calculation of the 
approximate solutions of the four-dimensional recurrence relation \eref{cost1}--\eref{cos3t2} 
obtained by inserting the expansion \eref{expansion} into these recurrence relations and identifying the different harmonics and powers of $\epsilon$. Bringing together different terms in \eref{expansion} we find
\numparts
\begin{eqnarray}
\label{an1_app}
a_n & = & \left(-1\right)^n \epsilon A_\epsilon\left(\xi\right)
+\epsilon^2 \widetilde{A}_2 \left(\xi \right)
+\epsilon^3 E_\epsilon\left(\xi\right),\\
b_{n+1} & = & \epsilon B_\epsilon\left(\xi\right)
+\left(-1\right)^n \epsilon^2 F_\epsilon\left(\xi\right),\\
c_n & = & \epsilon C_\epsilon\left(\xi\right)
+\left(-1\right)^n \epsilon^2 \widetilde{C}_2 \left(\xi \right)
+\left(-1\right)^n \epsilon^3 G_\epsilon\left(\xi\right),\\
\label{an4_app}
d_{n+1} & = & \left(-1\right)^n \epsilon^2 D_\epsilon\left(\xi\right)
+\epsilon^2 \widetilde{D}_2 \left(\xi \right)
+\epsilon^3 H_\epsilon\left(\xi\right).
\end{eqnarray}
\endnumparts
The functions $A_\epsilon\left(\xi\right)$, $B_\epsilon\left(\xi\right)$, $C_\epsilon\left(\xi\right)$, $D_\epsilon\left(\xi\right)$, $E_\epsilon\left(\xi\right)$, 
$F_\epsilon\left(\xi\right)$,
$G_\epsilon\left(\xi\right)$, $H_\epsilon\left(\xi\right)$ are of $\Or \left(1\right)$ as $\epsilon \rightarrow 0$. We have in addition
\begin{equation}
\label{eqc_app}
B_\epsilon \left(\xi \right)=-8C_\epsilon \left(\xi\right)+\Or \left(\epsilon \right).
\end{equation}
In the following, we shall omit the $\epsilon$-dependency in the notation. Inserting Ansatz \eref{an1_app}--\eref{an4_app} into the recurrence relations \eref{cost1}--\eref{cos3t2}, one obtains $\widetilde{D}_2=\widetilde{C}_2= \widetilde{A}_2 =0$, $F_\epsilon =\Or \left(\epsilon^2\right)$ and the equations
\numparts
\begin{eqnarray}
\fl 0=\tfrac{3}{2}\beta \epsilon^3 \left(-1\right)^{n+1} A\left(\xi\right)
\left[A^2\left(\xi\right) + \left(-1\right)^{n+1} A\left(\xi\right) \left(B\left(\xi\right)-C\left(\xi\right)\right)+2\left(B\left(\xi\right)-C\left(\xi\right)\right)^2\right]\nonumber\\
+\left(-1\right)^n \epsilon^2 \left(D\left(\xi\right) -D\left(\xi -\epsilon\right)\right)+ \epsilon^3 \left(-1\right)^n A\left(\xi\right)+ 2\epsilon^3 H\left(\xi \right)+\Or(\epsilon^4),\label{e8_app}\\
\fl 0=2\beta \epsilon^3 \left[
\tfrac{1}{4}\left(-1\right)^{n+1}A^3\left(\xi\right) + \tfrac{3}{2}A^2\left(\xi\right) \left(B\left(\xi\right)-C\left(\xi\right)\right)+\tfrac{3}{4}\left(B\left(\xi\right)-C\left(\xi\right)\right)^3
\right]\nonumber\\
+\epsilon \left(B\left(\xi\right) \!+\!B\left(\xi\! -\!\epsilon\right)\right)\!+\!\left(9\epsilon^2 \!+\!16\right)\epsilon C\left(\xi \right)\!+\!16\left(-1\right)^n\epsilon^3 G\left(\xi \right)\!+\Or (\epsilon^4),\label{e9_app}\\
\fl 0=\tfrac{7}{4}\left(-1\right)^{n+1} \epsilon^2 D\left(\xi\right)-\tfrac{7}{4}\epsilon^3 H\left(\xi\right)+ \left(-1\right)^{n+1} \epsilon \left(A\left(\xi +\epsilon \right) -A\left(\xi \right)\right)\nonumber\\
+2\epsilon^3 E\left(\xi \right)-\tfrac{3}{2}\beta \epsilon^3 A^2\left(\xi\right) \left(B\left(\xi\right)-C\left(\xi\right)\right)+\Or (\epsilon^4),\label{e10_app}\\
\fl 0=\left(9 \kappa \epsilon^2 +\tfrac{1}{4}\right)\epsilon B\left(\xi\right)
+ \epsilon \left(C\left(\xi +\epsilon \right) +C\left(\xi \right)\right)\nonumber\\
-{3}\beta \epsilon^3
\left[ A^2\left(\xi\right) \left(B\left(\xi\right)-C\left(\xi\right)\right)
+\tfrac{1}{2}\left(B\left(\xi\right)-C\left(\xi\right)\right)^3
\right]+\Or (\epsilon^4).\label{e11_app}
\end{eqnarray}
\endnumparts
Using \eref{e9_app} we get 
\begin{equation}
\label{eqg_app}
G\left(\xi \right)=\tfrac{1}{32}\beta A^3\left(\xi\right)+\Or (\epsilon).
\end{equation}
Now \eref{e9_app} allows to eliminate $C$ as a function of $A,B$ up to order $\epsilon^3$. Substituting the result into \eref{e11_app} yields
\begin{eqnarray}\label{eqb_app}
0&=&B\left(\xi +\epsilon \right)-2B\left(\xi\right)+B\left(\xi -\epsilon \right)
-9\epsilon^2 \left(\tfrac{1}{4}+16\kappa \right)B\left(\xi \right)\nonumber\\
&&+\tfrac{243}{8}\beta \epsilon^2 \left[
2A^2\left(\xi\right) B\left(\xi\right)
+\tfrac{81}{64}B^3\left(\xi\right)
\right]
+\Or (\epsilon^3).
\end{eqnarray}
Similarly, \eref{e8_app} and \eref{e10_app} lead to 
\begin{eqnarray}\label{eqh_app}
H\left(\xi \right)&=-\tfrac{3}{4}\beta A^2\left(\xi\right)\left(B\left(\xi\right)-C\left(\xi\right)\right)+\Or (\epsilon )\nonumber\\
&=-\tfrac{27}{32}\beta A^2\left(\xi\right)B\left(\xi\right)+\Or (\epsilon ),\\
\label{eqe_app}
E\left(\xi \right)&=\tfrac{7}{8}H\left(\xi\right)+\tfrac{3}{4}\beta A^2\left(\xi\right)\left(B\left(\xi\right)-C\left(\xi\right)\right)+\Or (\epsilon )\nonumber\\
&=\tfrac{27}{256}\beta A^2\left(\xi\right)B\left(\xi\right)+\Or (\epsilon ),\\
\label{eqd_app}
\epsilon D\left(\xi\right)&=-\tfrac{4}{7}\left(A\left(\xi+\epsilon\right)-A\left(\xi\right)\right)+\Or (\epsilon^3).
\end{eqnarray}
Using \eref{eqd_app} and \eref{eqc_app} in \eref{e8_app} yields
\begin{equation}\label{eqa_app}
\fl 0=A\left(\xi +\epsilon \right)-2A\left(\xi\right)+A\left(\xi -\epsilon \right)-\tfrac{7}{4}\epsilon^2 A\left(\xi \right)
+\tfrac{21}{8}\beta \epsilon^2 \left[
A^3\left(\xi\right)
+\tfrac{81}{32}A\left(\xi\right)B^2\left(\xi\right)
\right]
+\Or (\epsilon^3).
\end{equation}

We proceed by expanding $A,B,C,D,E,G,H$ up to first order in $\epsilon$, while $\xi$ is kept fix. Recalling that $A\left(\xi \right)=A_0\left(\xi\right) +\Or \left(\epsilon \right)$, $B\left(\xi \right)=B_0\left(\xi\right) +\Or \left(\epsilon \right)$ and using the same notation for the other variables, one obtains from the above equations as $\epsilon \rightarrow 0$
\numparts
\begin{eqnarray}\label{eqal_app}
0=\frac{\rmd^2A_0}{\rmd\xi^2}
-\tfrac{7}{4}A_0
+\tfrac{21}{8}\beta \left(
A^3_0
+\tfrac{81}{32}A_0B^2_0
\right),\\
\label{eqbl_app}
0=\frac{\rmd^2B_0}{\rmd\xi^2}
-9\left(\tfrac{1}{4}+16\kappa \right)B_0
+\tfrac{243}{8}\beta \left(
2A^2_0 B_0
+\tfrac{81}{64}B^3_0
\right),\\
\label{eqcl_app}
C_0=-\vphantom{\frac{1}{8}}\tfrac{1}{8}B_0,\\
\label{eqdl_app}
D_0=-\tfrac{4}{7}\frac{\rmd A_0}{\rmd\xi},\\
\label{eqel_app}
E_0=\vphantom{\frac{1}{8}}\tfrac{27}{256}\beta A^2_0B_0,\\
\label{eqgl_app}
G_0=\vphantom{\frac{1}{8}}\tfrac{1}{32}\beta A^3_0,\\
\label{eqhl_app}
H_0=-\vphantom{\frac{1}{8}}\tfrac{27}{32}\beta A^2_0B_0.
\end{eqnarray}
\endnumparts
Going back to \eref{an1_app}--\eref{an4_app}, we end up with
\numparts
\begin{eqnarray}
\label{an1l_app}
a_n & = & \left(-1\right)^n \epsilon A_0\left(\epsilon n\right)+\Or(\epsilon^2),\\
b_{n} & = & \epsilon B_0\left(\epsilon n\right)+\Or(\epsilon^2),\\
c_n & = & -\tfrac{1}{8}\epsilon B_0\left(\epsilon n\right)+\Or(\epsilon^2),\\
\label{an4l_app}
d_{n} & = & \tfrac{4}{7}\left(-1\right)^{n} \epsilon^2 \frac{\rmd A_0}{\rmd\xi}\left(\epsilon n\right)
+\Or(\epsilon^3),
\end{eqnarray}
\endnumparts
where $A_0,B_0$ are solutions of system \eref{eqal_app}--\eref{eqbl_app}. In this way we have obtained 
a formal continuum limit of the four-dimensional recurrence relation \eref{cost1}--\eref{cos3t2}
which provides approximate solutions valid for $\mu,\eta\approx0$.\\


\begin{thebibliography}{99}
\bibitem{MKAub} MacKay R S and Aubry S 1994 Proof of existence of breathers for time-reversible or Hamiltonian networks of weakly coupled oscillators {\em Nonlinearity} {\bf 7} 1623--1643
\bibitem{Bambusi} Bambusi D 1996 Exponential stability of breathers in Hamiltonian networks of weakly coupled oscillators {\em Nonlinearity} {\bf 9} 433--457
\bibitem{LiSpiMK} Livi R, Spicci M and MacKay R S 1997 Breathers on a diatomic FPU chain {\em Nonlinearity} {\bf 10} 1421--1434
\bibitem{AuKoKa} Aubry S, Kopidakis G and Kadelburg V 2001 Variational proof for hard discrete breathers in some classes of Hamiltonian dynamical systems {\em Discrete Contin.\ Dynam.\ Systems B} {\bf 1} 271--298
\bibitem{James} James G 2001 Existence of breathers on FPU lattices {\em C.\ R.\ Acad.\ Sci.\ Paris S\'er.\ I Math.} {\bf 332} 581--586
\bibitem{james2} James G 2003 Centre manifold reduction for quasilinear discrete systems {\em J.\ Nonlinear Sci.} {\bf 13} 27--63
\bibitem{JaNo} James G and Noble P 2003 Breathers on diatomic FPU chains with arbitrary masses {\em proceedings of the third conference on Localization and Energy Transfer in Nonlinear Systems, San Lorenzo de El Escorial, Spain, 2002} ed L Vazquez, R S MacKay and M P Zorzano (World Scientific)
\bibitem{jamesnoble} James G and Noble P 2004 Breathers on diatomic Fermi-Pasta-Ulam lattices {\em Physica D} {\bf 196} 124--171
\bibitem{Swanson_ea} Swanson B I, Brozik J A, Love S P, Strouse G F, Shreve A P, Bishop A R, Wang W-Z and Salkola M I 1999 Observation of intrinsically localized modes in a discrete low-dimensional material {\em Phys.\ Rev.\ Lett.} {\bf 82} 3288--3291
\bibitem{SchwarzEnSie} Schwarz U T, English L Q and Sievers A J 1999 Experimental generation and observation of intrinsic localized spin wave modes in an antiferromagnet {\em Phys.\ Rev.\ Lett.} {\bf 83} 223--226
\bibitem{BiUs} Binder P and Ustinov A V 2002 Exploration of a rich variety of breather modes in Josephson ladders {\em Phys.\ Rev.\ E} {\bf 66} 016603(4)
\bibitem{EdHamm} Edler J and Hamm P 2002 Self-trapping of the amide I band in a peptide model crystal {\em J.\ Chem.\ Phys.} {\bf 117} 2415--2424
\bibitem{Sato_ea} Sato M, Hubbard E, English L Q, Sievers A J, Ilic B, Czaplewski D A and Craighead H G 2003 Study of intrinsic localized vibrational modes in micromechanical oscillator arrays {\em Chaos} {\bf 13} 702--715
\bibitem{Mandelik_ea} Mandelik D, Eisenberg H, Silberberg Y, Morandotti R and Aitchison J S 2003 Observation of mutually-trapped multi-band optical breathers in waveguide arrays {\em Phys.\ Rev.\ Lett.} {\bf 90} 253902(4)
\bibitem{FlaWi} Flach S and Willis C R 1998 Discrete breathers {\em Phys.\ Rep.} {\bf 295} 181--264
\bibitem{AlfBraKo} Alfimov G L, Brazhnyi V A and Konotop V V 2004 On classification of intrinsic localized modes for the discrete nonlinear Schrödinger equation {\em Physica D} {\bf 194} 127--150
\bibitem{FePaU} Fermi E, Pasta J and Ulam S 1955 Studies of nonlinear problems. I. {\em Technical Report LA-1940, Los Alamos National Laboratory}
\bibitem{BuKiPy} Burlakov V M, Kiselev S A and Pyrkov V N 1990 Computer simulations of intrinsic localized modes in 1-D anharmonic lattices {\em Solid State Commun.} {\bf 74} 327--331
\bibitem{Aoki} Aoki M 1992 Self-localized mode in a diatomic nonlinear lattice {\em J.\ Phys.\ Soc.\ Japan} {\bf 61} 3024--3026
\bibitem{Flach} Flach S 1994 Conditions on the existence of localized excitations in nonlinear discrete systems {\em Phys.\ Rev.\ E} {\bf 50} 3134--3142
\bibitem{CreLiSpi} Cretegny T, Livi R and Spicci M 1998 Breather dynamics in diatomic FPU chains {\em Physica D} {\bf 119} 88--98
\bibitem{MaAub} Mar\'{\i}n J L and Aubry S 1996 Breathers in nonlinear lattices: numerical calculation from the anticontinuous limit {\em Nonlinearity} {\bf 9} 1501--1528
\bibitem{MacSep} Sepulchre J-A and MacKay R S 1997 Localized oscillations in conservative or dissipative networks of weakly coupled autonomous oscillators {\em Nonlinearity} {\bf 10} 679--713
\bibitem{SieTa} Sievers A J and Takeno S 1988 Intrinsic localized modes in anharmonic crystals {\em Phys.\ Rev.\ Lett.} {\bf 61} 970--973
\bibitem{franc1} Franchini A, Bortolani V and Wallis R F 1996 Intrinsic localized modes in the bulk and at the surface of anharmonic diatomic chains {\em Phys.\ Rev.\ B} {\bf 53} 5420--5429
\bibitem{franc2} Franchini A, Bortolani V and Wallis R F 2002 Theory of intrinsic localized modes in diatomic chains: beyond the rotating wave approximation {\em J.\ Phys.: Condens.\ Matter} {\bf 14} 145--152
\bibitem{Konotop} Konotop V V 1996 Small-amplitude envelope solitons in nonlinear lattices {\em Phys.\ Rev.\ E} {\bf 53} 2843--2858
\bibitem{nikos} Flytzanis N, Malomed B A and Neuper A 1998 Odd and even intrinsic modes in a diatomic nonlinear lattice {\em Physica D} {\bf 113} 191--195
\bibitem{gian3} Giannoulis J and Mielke A 2006 Dispersive evolution of pulses in oscillator chains with general interaction potentials {\em Discrete Contin.\ Dynam.\ Systems B} {\bf 6} 493--523
\bibitem{yang} Yang J 1997 Classification of the solitary waves in coupled nonlinear Schr\"odinger equations {\em Physica D} {\bf 108} 92--112
\bibitem{boyd} Boyd J P and Tan B 1999 Composite bound states of wide and narrow envelope solitons in the coupled Schr\"odinger equations through matched asymptotic expansions {\em Nonlinearity} {\bf 12} 1449--1469
\bibitem{champneys} Champneys A R and Yang J 2002 A scalar nonlocal bifurcation of solitary waves for coupled nonlinear Schr\"odinger systems {\em Nonlinearity} {\bf 15} 2165--2192
\bibitem{ioossjos} Iooss G and Joseph D D 1990 {\em Elementary stability and bifurcation theory} (Springer).
\end{thebibliography}
\end{document}